\documentclass[aps,prd,twocolumn,groupedaddress,nofootinbib,superscriptaddress]{revtex4-1}  
\usepackage{graphics,mathtools,tabu}
\usepackage{epstopdf}
\usepackage{amsmath}
\usepackage{amsfonts}
\usepackage[colorlinks=true,citecolor=red,linkcolor=blue,breaklinks=true]{hyperref}
\usepackage[figure,table]{hypcap}
\usepackage{amssymb}
\usepackage{appendix}
\usepackage{comment}
\usepackage{bbold}
\usepackage{color}
\usepackage{slashed}
\usepackage{subfigure}
\usepackage{setspace}
\usepackage{footnote}
\usepackage{multirow}
\usepackage{braket}
\usepackage[normalem]{ulem}
\usepackage[a4paper, portrait, margin=0.5in]{geometry}
\usepackage{mathrsfs}
\usepackage{longtable}
\usepackage{enumitem}

\bibpunct{[}{]}{,}{n}{}{,}
\newcommand{\mchi}{m_{\chi}}
\renewcommand{\eqref}[1]{Eq.\ (\ref{#1})}
\newcommand{\eqsref}[2]{Eqs.\ (\ref{#1})--(\ref{#2})}
\newcommand{\upq}{U(1)_{\rm PQ}}
\newcommand{\sigchiN}{\sigma_{\chi N}}
\newcommand{\chionevev}{\chi_{1_{\rm min}}}
\newcommand{\chitwovev}{\chi_{2_{\rm min}}}

\begin{document}
\singlespacing
\preprint{\\~\\TIFR/TH/18-31\\NUHEP-TH/18-11\\HRI-RECAPP-2018-013}

\title{Mixed WIMP-axion dark matter}

\author{Suman Chatterjee}
\email{suman.chatterjee@tifr.res.in}
\affiliation{Tata Institute of Fundamental Research, Homi Bhabha Road, Mumbai 400\,005, India}
\author{Anirban Das}
\email{anirbandas@theory.tifr.res.in}
\affiliation{Tata Institute of Fundamental Research, Homi Bhabha Road, Mumbai 400\,005, India}
\author{Tousik Samui}
\email{tousiksamui@hri.res.in}
\affiliation{Regional Centre for Accelerator-based Particle Physics, Harish-Chandra Research Institute, HBNI, Chhatnag Road, Jhunsi, Allahabad, 211\,019, India}
\author{Manibrata Sen}
\email{manibrata@berkeley.edu}
\affiliation{Department of Physics and Astronomy, Northwestern University, 2145 Sheridan Road, Evanston, Illinois 60208, USA}
\affiliation{Department of Physics, University of California Berkeley, Berkeley, California 94720, USA}

\begin{abstract}
We study the experimental constraints on a model of a
two-component dark matter, consisting of the QCD axion, and
a scalar particle, both contributing to the dark matter
relic abundance of the Universe. The global Peccei-Quinn
symmetry of the theory can be spontaneously broken down to a
residual $\mathbb{Z}_2$ symmetry, thereby identifying this
scalar as a stable weakly interacting massive particle,
i.e., a dark matter candidate, in addition to the axion. We
perform a comprehensive study of the model using the latest
data from dark matter direct and indirect detection
experiments, as well as new physics searches at the Large
Hadron Collider. We find that although the model is mostly
constrained by the dark matter detection experiments, it is
still viable around a small region of the parameter space
where the scalar dark matter is half as heavy as the
Standard Model Higgs. In this allowed region, the bounds
from these experiments are evaded due to a cancellation
mechanism in the dark matter--Higgs coupling. The collider
search results, however, are shown to impose weak bounds on
the model.
\end{abstract}

\maketitle
\section{Introduction}\label{sec:intro}
The evidence of \emph{cold dark matter} (CDM) is
overwhelming from the cosmological data, even though its
detection and  identification continue to be one of the most
interesting and challenging problems
today\,\cite{Ade:2015xua}. Many particle dark matter (DM)
models have been proposed over the last few decades, one of
the oldest of them being the weakly interacting massive
particle (WIMP)
model\,\cite{Jungman:1995df, Pagels:1981ke,Kolb:1983fm,Ellis:1983ew}
(for reviews,
see\,\cite{Bertone:2004pz,Bergstrom:2000pn,Bertone:2016nfn}).
In the WIMP scenario, the dark matter relic abundance is
obtained through the annihilation of dark matter particles
in the early Universe with weak scale cross sections and
electroweak scale
masses\,\cite{Jungman:1995df,Gondolo:1990dk,Griest:1990kh,Steigman:2012nb}.
The fact that one gets new physics at the electroweak scale
for a WIMP mass $\sim\!100{\rm\ GeV}$ makes this scenario a
very appealing solution to the dark matter
problem\,\cite{Steigman:1984ac}.

The absence of $CP$ violation in the strong sector of the
Standard Model (SM) is another long-standing puzzle in the
particle physics community\,\cite{Peccei:2006as}. The null
results of the neutron electric dipole moment measurement
experiments so far restrict the value of the coefficient
$\theta_{\rm QCD}$ of the parity-violating
$\mathbf{E}\cdot\mathbf{B}$ operator to be less than
$10^{-10}$\,\cite{Crewther:1979pi}. In the present form of
the SM, this is a fine-tuning problem since there is no
symmetry that protects such a small number from large
higher-order corrections\,\cite{Kim:2008hd}. Therefore, a
natural explanation of the smallness of strong $CP$
violation is sought, and an elegant solution to this puzzle
is given by the introduction of a global $U(1)$ Peccei-Quinn
(PQ)
symmetry\,\cite{Peccei:1977hh,Wilczek:1977pj,Weinberg:1977ma,Kim:1979if,Shifman:1979if}.
This symmetry is spontaneously broken at a scale much larger
than the electroweak scale by a scalar field, with the axion
as the corresponding massless Nambu-Goldstone boson of this
$\upq$ symmetry. The coefficient $\theta_{\rm QCD}$ is
dynamic in this model and its small value is naturally
attained in this way and is inversely proportional to the PQ
scale. In this context, a large number of axion models have
been proposed in the literature. The early PQ model, first
proposed in \cite{Peccei:1977hh} and further developed in
\cite{Wilczek:1977pj,Weinberg:1977ma}, augments the SM with
an additional complex scalar, charged under the electroweak
(EW) symmetry. The Lagrangian is additionally invariant
under a global $U(1)$ symmetry, which is spontaneously
broken at the EW scale. However, this model predicts large
axion couplings and hence is ruled out by
experiments\,\cite{Kim:1986ax}. To circumvent the
experimental bounds, invisible axion models were proposed
independently: the Kim-Shifman-Vainshtein-Zakharov (KSVZ)
model\,\cite{Kim:1979if,Shifman:1979if} and the
Dine-Fischler-Srednicki-Zhitnitsky
(DFSZ)\,\cite{Dine:1981rt,Zhitnitsky:1980tq} model. The KSVZ
model introduces heavy, colored, EW singlet quarks, in
addition to the PQ scalar. In this model, the axions have no
direct tree-level couplings with the SM fields and can only
have induced couplings to the SM leptons. On the other hand,
the DFSZ model introduces an additional Higgs field, along
with the PQ scalar. This allows the axions to have natural
couplings to leptons at tree level. Other than this,
axionlike Goldstone particles arise in a multitude of other
theoretical scenarios as well, e.g.,
majorons\,\cite{Chikashige:1980ui,Gelmini:1980re},
familons\,\cite{Wilczek:1982rv,Berezhiani:1990wn,Jaeckel:2013uva},
axions from string
theory\,\cite{Witten:1984dg,Conlon:2006tq,Cicoli:2012sz},
axions from accidental symmetry
breaking\,\cite{Georgi:1981pu,Dias:2014osa,Choi:2009jt},
etc.

The axion field gains a small mass inversely proportional to
the $\upq$-breaking scale, after the QCD condensation at a
temperature of about $T\simeq 200$ MeV. In the early
Universe, the axion can be produced nonrelativistically
through a coherent oscillation of the axion field due to the
misalignment of the PQ vacuum. This is known as the
\emph{misalignment mechanism} of axion
production\,\cite{Kuster:2008zz,Marsh:2015xka}. The axion is
not completely stable; however, it has very feeble couplings
with SM particles, thereby ensuring a lifetime longer than
the age of the Universe\,\cite{Kim:2007qa}. This makes the
axion a very good CDM
candidate\,\cite{Preskill:1982cy,Abbott:1982af,Dine:1982ah},
although the same feeble couplings make direct detection of
these axions challenging\,\cite{Abbott:1982af}.

Since current DM detection experiments have shown null
results, one needs to look for other alternatives to the
simple WIMP scenario. One such alternative is a
multicomponent DM, where one component acts as a WIMP,
whereas the other components might have very different
interactions. These models are less constrained due to the
fact that the fraction of WIMPs in these multicomponent DM
has not been determined experimentally and hence is a free
parameter. This has been shown to have interesting
phenomenological
consequences\,\cite{Zurek:2008qg,Biswas:2013nn,Bhattacharya:2013hva,DuttaBanik:2016jzv,Arcadi:2016kmk, Alves:2016bib,Pandey:2017quk,Chakraborti:2018aae}.
 
In this work, we study a two-component DM model consisting
of a WIMP and the axion as the DM candidates. This type of
model gives a unifying scenario where the PQ field, which is
motivated to solve the strong-$CP$ problem, and the WIMP,
which is a natural solution to DM puzzle, can be
accommodated in a single
go\,\cite{Dasgupta:2013cwa,Alves:2016bib}. Furthermore,
these models can also be extended to include neutrino
masses\,\cite{Dasgupta:2013cwa}. Hence, these models, and
their variations thereof, can account for three of the most
important puzzles in the SM. Possible UV completions are
considered in\,\cite{Alves:2016bib}. As a simple realization
of this, one can consider the KSVZ model of axion with an
additional scalar field charged under the
$U(1)_{\rm PQ}$\,\cite{Dasgupta:2013cwa}. This additional
scalar gets its stability from the residual $\mathbb{Z}_2$
symmetry of the broken $\upq$ and hence becomes a WIMP-like
DM candidate\,\cite{PhysRevLett.62.1221}. Breaking of the
$\upq$ and the electroweak symmetry leads to a mixing
between the Higgs and the radial part of the PQ scalar,
which leads to interesting phenomenological consequences.
The advantage of this model is that although the axions have
very weak interactions with the SM, the coupling between
this dark scalar and the SM Higgs doublet provides a portal
to test this model in different DM detection experiments,
both direct and indirect. The model can also give different
signatures at collider experiments. For example, the KSVZ
model predicts new colored, electroweak singlet quarks,
which can be produced at colliders. Mixing with a scalar
affects the properties of the Higgs boson, which can be
directly used to constrain the mixing parameters.
Furthermore, the dark scalar can also contribute to momentum
imbalance in a collision event.

Hence, in the light of recent experiments, we explore the
constraints on the WIMP-axion DM model, both from DM search
experiments as well as collider searches. Using the latest
limit on DM-nucleon scattering cross section from
XENON1T$\times$1 yr experiment data\,\cite{Aprile:2018dbl},
we find that the phenomenologically interesting mass range
of $m_{\rm DM}\gtrsim 100{\rm\ GeV}$ is ruled out in such
models. However, the stringent bounds from XENON1T$\times$1
yr data can be evaded in a small region of the parameter
space where the scalar dark matter is half as heavy as the
Higgs. This is a direct outcome of the mixing of the Higgs
with the scalar, which leads to a cancellation mechanism in
the Higgs portal coupling, thereby reducing the DM-nucleon
scattering cross section. As a result, while minimal scalar
DM models are mostly ruled out by direct detection
bounds\,\cite{Athron:2017kgt}, such WIMP-axion models can
still survive with a reduced parameter space. Collider
signals, on the other hand, are highly plagued by the
backgrounds from the production of Standard Model particles,
and hence the signals are not significant enough to be
observed above the
background\,\cite{Aaboud:2016tnv,Sirunyan:2017hci,Sirunyan:2017jix}.

The paper is organized as follows. Section~\ref{sec:model}
discusses the model and the different parameters involved.
Section~\ref{sec:expt-bound} talks about the different
experimental bounds, and how they constrain the parameters
of the model. In Sec.~\ref{sec:result}, we summarize the
main results, and finally in Sec.~\ref{sec:conclude}, we
conclude.

\section{The Model}\label{sec:model}
We consider the KSVZ model of the axion, where electroweak
singlet quarks $Q_L$ and $Q_R$ and a complex scalar $\zeta$,
both transforming under a global $U(1)_{\rm PQ}$ symmetry,
are added to the SM\,\cite{Kim:1979if,Shifman:1979if}. These
quarks are vectorlike and hence do not introduce any chiral
anomaly\,\cite{Adler:1969gk,Bell:1969ts}. We augment this
model with a complex scalar
$\chi\!=\!(\chi_1+i \chi_2)/\sqrt{2}$ which is a SM singlet
but charged under the $\upq$
symmetry\,\cite{Dasgupta:2013cwa}. The axion $a$ is the
Nambu-Goldstone mode of the scalar field $\zeta$, which can
couple to the vectorlike quarks as well as $\chi$. As in the
original KSVZ model, the axion can act as a CDM
candidate\,\cite{Abbott:1982af}. The charges and quantum
numbers of the new particles are listed in
Table~\ref{tab:charges}.

\begin{ruledtabular}
\begin{table}[h]
\caption{\small New particles in the model and their
charges. PQ charges of all the SM particles are zero.}
\label{tab:charges}
\begin{tabular}{lcccc}
 &&&&\\[-2.5ex] & $\zeta$ & $\chi$ & $Q_L$  & $Q_R$ \\[0.5ex]
\hline &&&&\\[-2ex] Spin         &    0    &    0   &  1/2   &  1/2   \\[0.5ex]
 &&&&\\[-2.5ex] $SU(3)_C$    &    1    &    1   &   3    &   3    \\[0.5ex]
 &&&&\\[-2.5ex] $SU(2)_L$    &    1    &    1   &   1    &   1    \\[0.5ex]
 &&&&\\[-2.5ex] $U(1)_Y$     &    0    &    0   & $-1/3$ & $-1/3$ \\[0.5ex]
 &&&&\\[-2.5ex] $U(1)_{PQ}$  &    2    &    1   &   1    &  $-1$  \\[0.5ex]
\end{tabular}
\end{table}
\end{ruledtabular}

The relevant part of the Lagrangian, governing the
interactions of $Q_{L,R},\,\zeta$, and $\chi$ with the SM,
is given by
\begin{eqnarray}
\mathcal{L} &\supset& -\lambda_H\left(|H|^2 - \frac{v_H^2}{2}\right)^2
                      -\lambda_\zeta\left(\left|\zeta\right|^2 -\frac{F_a^2}{2}\right)^2 \nonumber\\
            &       & -\lambda_{\zeta H} \left(\left|H\right|^2 - \frac{v_H^2}{2}\right) \left(\left|\zeta\right|^2 -\frac{F_a^2}{2}\right) - \lambda_\chi \left|\chi\right|^4 \nonumber\\
            &       & -\mu_\chi^2 \left|\chi\right|^2- \lambda_{\chi H} \left|H\right|^2 \left|\chi\right|^2
                - \lambda_{\zeta\chi} \left|\zeta\right|^2 \left|\chi\right|^2 \nonumber\\
            &       & + \Big[\epsilon_\chi \zeta^* \chi^2 
                + f_d \chi \bar{Q}_L d_R
                + f_Q \zeta \bar{Q}_L Q_R + \text{H.c.} \Big]\,.\label{eqn:lagrangian}
\end{eqnarray}
Here $H$ is the SM Higgs doublet and $d_R$ represents
right-handed down-type quarks in the SM. After electroweak
symmetry breaking via the Higgs vacuum expectation value
(VEV) $v_H$, one has $|H|=(h_0+v_H)/\sqrt{2}$, where $h_0$
is the Higgs boson. Similarly, using the nonlinear
representation, one can write
$\zeta=e^{ia\,/F_a}\left(F_a+\sigma_0\right)/\sqrt{2}$,
where $F_a$ is the $\upq$-symmetry-breaking scale, also
known as the axion decay constant, and $\sigma_0$ is the
radial excitation of the $\zeta$ field. Constraints from
supernova cooling data disfavor values of $F_a$ smaller than
$10^{10}{\rm\ GeV}$\,\cite{Raffelt:1987yt}.

After the breaking of both the symmetries, viz. electroweak
and PQ symmetries, the interaction term between $H$ and
$\zeta$ fields leads to mixing between $h_0$ and $\sigma_0$
with the mass matrix
\begin{equation}
    M^2 \equiv
    \begin{pmatrix}
        2v_H^2\lambda_H           && F_a v_H \lambda_{\zeta H}\\
        F_a v_H \lambda_{\zeta H} && 2 F_a^2 \lambda_\zeta\\
    \end{pmatrix}. \label{eqn:mass-matrix}
\end{equation}
As a result of the mixing, the scalars in the mass basis are
related to those in the flavor basis as
\begin{equation}\label{massbasis}
\begin{pmatrix}
    h \\ \sigma
\end{pmatrix}
    =
\begin{pmatrix}
    \cos\,\theta && -\sin\,\theta\\
    \sin\,\theta &&  \cos\,\theta\\
\end{pmatrix}
\begin{pmatrix}
    h_0\\ \sigma_0
\end{pmatrix},
\end{equation}
where the mixing angle, in the limit $F_a\gg v_H$, is given
by
\begin{equation}
\sin\,\theta \simeq \frac{v_H}{F_a}\frac{\lambda_{\zeta H}}{2\lambda_\zeta}.\label{eqn:sintheta}
\end{equation}
One obtains the masses of the physical states as
\begin{eqnarray}
m_h      &\simeq&\,v_H\sqrt{2\lambda_H\left(1-\frac{\lambda_{\zeta H}^2}{4\lambda_H\lambda_\zeta}\right)}\, + \mathcal{O}\left(\frac{v_H}{F_a}\right),\label{eqn:mh}\\
m_\sigma &\simeq&\,F_a\sqrt{2\lambda_\zeta}+\mathcal{O}\left(\frac{v_H}{F_a}\right).     \label{eqn:msig}
\end{eqnarray}
Note that the mass $m_h$ of the mixed state $h$ is no longer
$\sqrt{2\lambda_H v_H^2}$, as predicted by the SM. Since $h$
is the physical state, we fix $m_h$ at $125$~GeV and the
Higgs VEV $v_H$ at $246$~GeV to match with the
experimentally measured masses of the observed
scalar\,\cite{Aad:2012tfa,Chatrchyan:2012xdj} and $W$ and
$Z$ bosons, respectively\,\cite{Tanabashi:2018oca}. The
value of $\lambda_H$ is no longer the SM value,
$\lambda^\text{SM}_H\simeq 0.13$, but is dependent on other
parameters in this model and can be calculated using
\eqref{eqn:mh}. In fact, if we take
$\lambda_H = \lambda_H^\text{SM} =\frac{m_h^2}{2 v_H^2}$,
from \eqref{eqn:mh} it is evident that $\lambda_{\zeta H}$
has to be zero; i.e., the SM Higgs does not mix with
$\zeta$, as considered in\,\cite{Dasgupta:2013cwa}. Note
that there is no underlying symmetry in the theory that
allows us to set $\lambda_{\zeta H}$ to zero in the
Lagrangian. More importantly, although the mixing is very
small, the relation between the masses of the physical
states with other model parameters plays a major role in
imposing constraints on the model. Therefore, we do not
neglect the mixing of $h_0$ with $\sigma_0$ in this study.

The masses of $\chi_1$ and $\chi_2$ are given by 
\begin{equation}\label{masschi}
    m_{\chi_{1,2}}^2 = \frac{1}{2}\left(2\mu_\chi^2+v_H^2\lambda_{\chi H}
                     +F_a^2\lambda_{\zeta\chi}\mp2\sqrt{2}F_a\epsilon_\chi\right).
\end{equation}
Without loss of generality, we can take $\epsilon_\chi\!>0$
such that $m_{\chi_{1}}\!<m_{\chi_{2}}$; hence $\chi_1$ can
be the DM candidate, and we, henceforth, denote the mass of
$\chi_1$ as just $\mchi$. Note that after the PQ-symmetry
breaking, the Lagrangian in \eqref{eqn:lagrangian} has a
residual $\mathbb{Z}_2$ symmetry which stabilizes $\chi_1$.
It should also be noted that in \eqref{masschi},
$\mu_\chi^2$ is defined to be negative and hence cancels out
the large contribution coming from $F_a$. This type of
fine-tuning is a general feature of these axion
models\,\cite{Giddings:1988cx,Coleman:1988tj,Kamionkowski:1992mf,Ghigna:1992iv,Dasgupta:2013cwa}.
Since the fine-tuning is required mainly in the dark sector,
we do not explore it further and defer the details to a
later work. Furthermore, one can also motivate a tiny value
of $\epsilon_\chi$ from naturalness arguments. As
$\epsilon_\chi\rightarrow 0$, one obtains an extra $U(1)$
symmetry in the theory, apart from the $U(1)_{\rm PQ}$. This
can allow $\epsilon_\chi$ to be naturally small. Note that a
small $\epsilon_\chi$ does not necessarily mean a small mass
split between the two DM states. For example, we shall
consider values around $F_a \sim 10^{10}$~GeV, which means
the mass split is
$\Delta_\chi\simeq\sqrt{F_a\epsilon_\chi} \sim 1{\rm\ TeV}$.
This is much larger than the mass of the lighter state.

At this point, it is important to note that in this setup
the complex scalar $\chi$ can, in principle, develop a VEV
before the PQ field $\zeta$. This can be prevented if
parameters are tuned such that the VEV of $\chi$,
$v_\chi = \sqrt{-\left(\mu_\chi^2/2\lambda_\chi\right)}$,
remains smaller than that of $\zeta$. It is always possible
to tune $\epsilon_\chi$ and $\mu_\chi$ such that the mass of
$\chi_1$ remains fixed, and $v_\chi$ remains below $F_a$.
This is because the only place where $\mu_\chi$ and $F_a$
appear is in the expression for the masses of the real
scalars in \eqref{masschi}. Since the mass difference
between $\chi_1$ and $\chi_2$ is proportional to
$\epsilon_\chi$, this has the effect of changing the mass of
the heavier scalar $\chi_2$. However, in our analysis,
$\chi_2$ is heavy enough to be decoupled from the particle
content. With this setup, one can ensure that at a very high
scale, around $F_a$, breaking of the PQ symmetry occurs
before the symmetry breaking of $\chi$. After the
PQ-symmetry breaking, $\chi_1$ and $\chi_2$ may or may not
remain tachyonic depending on the choice of parameters.
However, we have to ensure that none of them can develop a
VEV earlier than the electroweak scale in order to make our
mechanism work. Once the Higgs develops a VEV, both $\chi_1$
and $\chi_2$ become nontachyonic, irrespective of whether
they were tachyonic or not prior to the electroweak symmetry
breaking. The details of this mechanism is worked out in the
Appendix.

The mass of the axion is generated through nonperturbative
QCD effects and is inversely proportional to
$F_a$\,\cite{Gross:1980br};
\begin{equation}
    m_a\simeq 0.6\,{\rm meV}\,\times \left(\frac{10^{10}\,{\rm GeV}}{F_a}\right).
\end{equation}
The couplings of the axion to SM particles are also
suppressed by inverse power of $F_a$, so the decay lifetime
of the axion is very large. In fact, if we take the value of
$F_a>10^{10}$~GeV, as allowed by the supernova cooling
data\,\cite{Raffelt:1987yt}, its lifetime becomes larger
than the age of the Universe. Thus, the axion also acts as a
viable candidate for CDM in this model. Therefore, both
$\chi_1$ and the axion will contribute to the total DM relic
density in the Universe.

Finally, the vectorlike quarks obtain their mass
$m_Q=f_Q F_a/\sqrt{2}$ after $\zeta$ develops a VEV. If this
mass is $\sim\mathcal{O}({\rm TeV})$, they can be produced
at the LHC. This is expected to give direct constraints on
this model; however, in order to have a mass of
$\sim\mathcal{O}({\rm TeV})$, the coupling $f_Q$ needs to be
extremely tiny $\sim\mathcal{O}(10^{-6})$.

The new interactions introduce two portals connecting the SM
and the dark sector through the Higgs (via the
$h\chi_1\chi_1$) and the down-type quark (via the
$\chi_1\bar{Q}_L d_R$). Of the two, the $h\chi_1\chi_1$
interaction is the more important one and will play a key
role in our analysis. The $h\chi_1\chi_1$ coupling is given
by
\begin{equation}
    g_{h\chi_1\chi_1} = i\left(F_a \lambda_{\zeta\chi}\sin\theta - v_H \lambda_{\chi H} \cos\theta - \sqrt{2}\epsilon_\chi \sin\theta\right). \label{eqn:hChiChi}
\end{equation}
Although $\sin\theta$ is small, the first term cannot be
ignored due to the large scale $F_a$. Using the
approximation for $\sin\theta$ in \eqref{eqn:sintheta}, we
obtain
\begin{equation}
    g_{h\chi_1\chi_1} \simeq i\, v_H \left(\frac{\lambda_{\zeta\chi}\lambda_{\zeta H}}{2\lambda_\zeta} - \lambda_{\chi H} \right). \label{eqn:approxhChiChi1}
\end{equation}
Note that in the presence of nonzero $\lambda_{\zeta H}$,
the $h\chi_1\chi_1$ coupling vanishes at
\begin{equation}\label{shift}
 \lambda_{\chi H} = \frac{\lambda_{\zeta\chi}\lambda_{\zeta H}}{2\lambda_\zeta}\,.
\end{equation}
This is where we differ from\,\cite{Dasgupta:2013cwa}, where
the authors had set $\lambda_{\zeta H}=0$, which led to the
vanishing of the $h\chi_1\chi_1$ coupling at
$\lambda_{\chi H} = 0$ . This shift will play a crucial role
in the following analysis.

\begin{figure}[h]
\includegraphics[width=0.4\textwidth]{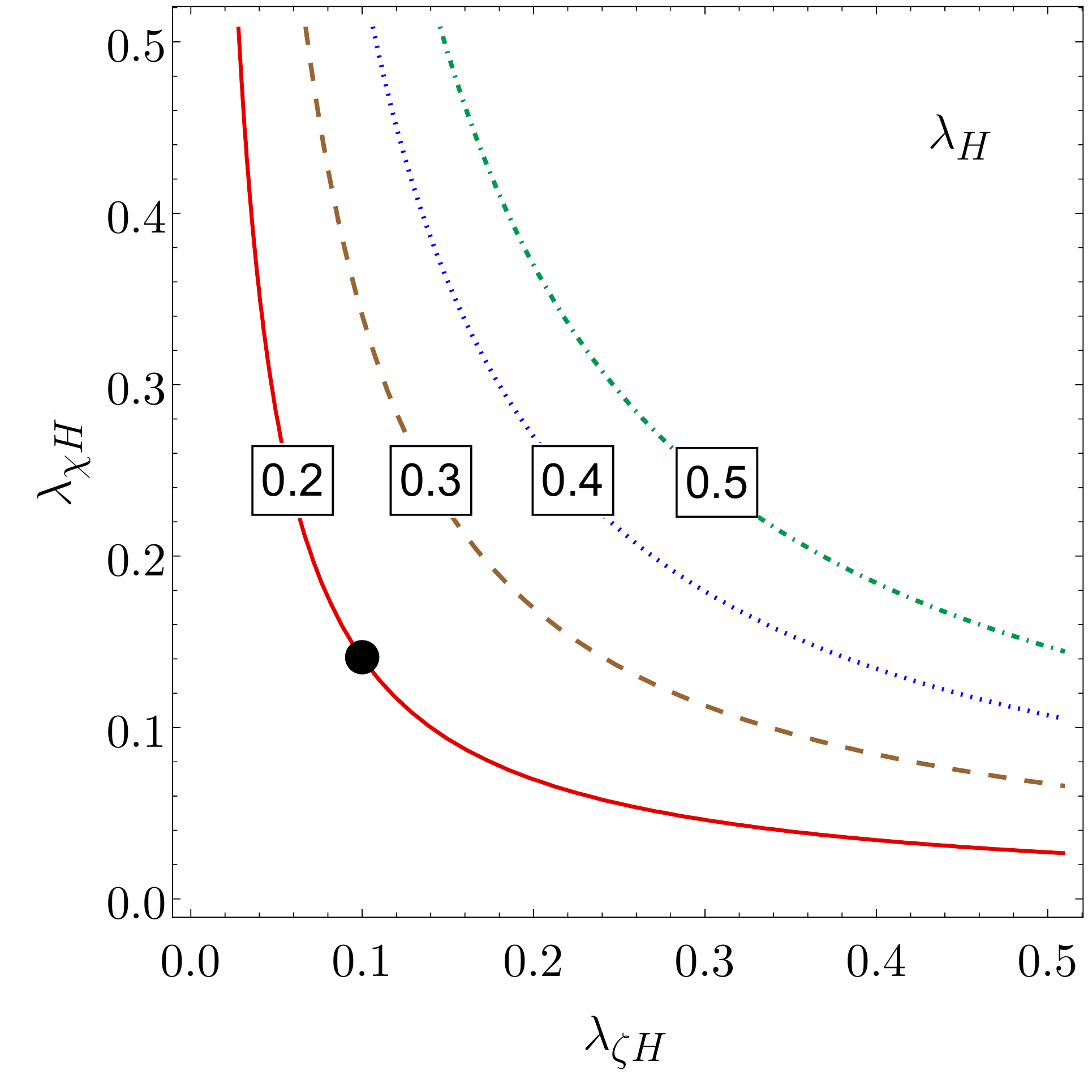}
\caption{\small Contours of $\lambda_H$, for which the
$h\chi_1\chi_1$ coupling vanishes. The other parameters
considered are
$\lambda_{\zeta\chi}=0.1,\, v_H = 246\,{\rm GeV}$ and
$m_h=125{\rm\ GeV}$. The benchmark point chosen for further
analysis, $\lambda_{\zeta H}=0.1,\,\lambda_{\chi H}=0.14$,
and $\lambda_H=0.2$, is shown as a black circle.}
\label{fig:lambdaH_contour}
\end{figure}

Using \eqref{eqn:mh}, $\lambda_\zeta$ can be written in
terms of $m_h,\, \lambda_{\zeta H}$, and $\lambda_H$. This
gives a family of solutions, satisfying \eqref{shift}. In
Fig.\,\ref{fig:lambdaH_contour}, we show four contours of
$\lambda_H$ in the $\lambda_{\zeta H}-\lambda_{\chi H}$
plane for a given value of $\lambda_{\zeta\chi}=0.1$. Any
point on these hyperbolas satisfies \eqref{shift}, leading
to vanishing $h\chi_1\chi_1$ coupling. The benchmark point
chosen for further analysis,
$\lambda_{\zeta H}=0.1,\,\lambda_{\chi H}=0.14$ and
$\lambda_H=0.2$, is shown as a black circle in the figure.
One can, in principle, probe other values of $\lambda_H$ in
this parameter space, but we do not show them here for
clarity. However, one should not take
$\lambda_H < \lambda_H^\text{SM} \simeq 0.13$ since it leads
to negative values of $\lambda_\zeta$, thereby making the
potential for  $\zeta$ unstable.

Finally, note from \eqref{eqn:msig} that the mass of
$\sigma$ is proportional to the $U(1)_{PQ}$-breaking scale
$F_a$. So if $\lambda_\zeta \sim \mathcal{O}(1)$, $\sigma$
becomes very heavy and decouples from the low-energy theory.
Therefore, for all practical purposes, $\sigma$ does not
play any significant role in present experiments. However,
it is possible to have the mass of $\sigma$ at around TeV,
but only within a highly fine-tuned region of the parameter
space.

One may wonder as to how much fine-tuning might be necessary
for this scenario. Without going into details, we provide a
back-of-the-envelope estimate here. From
\eqref{eqn:mass-matrix}, if $\lambda_\zeta\!\sim\!10^{-14}$,
then both the scalars $h$ and $\sigma$ can have a mass
$\sim\!\mathcal{O}(100)$~GeV. However, in order to keep the
physical masses real, i.e., both the eigenvalues of the mass
matrix positive, the off-diagonal terms have to be of the
same order as the diagonal terms. This requires
$\lambda_{\zeta H}$ to be further fine-tuned to values
$\sim10^{-7}$. However, such small values of $\lambda_\zeta$
and $\lambda_{\zeta H}$ will raise the value of
$g_{h\chi_1\chi_1}$ [see Eqs.~(\ref{eqn:hChiChi}) and
(\ref{eqn:approxhChiChi1})] to values $\gg 1$, which makes
the whole problem highly nonperturbative. Then, one would
again need to choose $\lambda_{\zeta\chi}$ unnaturally small
to solve this issue.\footnote{Note that the results given in
\eqsref{eqn:mh}{eqn:approxhChiChi1} were obtained in the
limit $F_a \gg v_H$.  This approximation breaks down when
the $\lambda$s are set to such small values. Hence one has
to start from the mass matrix in \eqref{eqn:mass-matrix}
and proceed without any approximation to arrive at this
conclusion.}

Since the above scenario is fine-tuned, we do not pursue it
here. Rather, we consider natural values of all couplings
$\lesssim\mathcal{O}(1)$. As a result, in this work, the
heavy scalar $\sigma$ decouples early on and does not enter
our analysis.

\section{Experimental Probes of Dark Matter}\label{sec:expt-bound}
Naturally, this model will have vast implications for dark
matter search experiments. In addition, the LHC search for
heavy vectorlike particles, as well as missing energy
searches, will also test this model. Using
{\tt FeynRules}\,\cite{Christensen:2008py,Alloul:2013bka} to
implement the model, we constrain it with the latest results
from these experiments. Broadly, three avenues are explored:
\begin{enumerate}
    \item DM relic density constraint, direct and indirect
    detection experiments set limits on the parameters
    connecting the dark sector with the visible sector.
    
    \item Mixing between $h_0$ and $\sigma_0$ changes the
    couplings of the observed 125~GeV scalar from that of
    the SM Higgs. This leads to changes in the properties of
    the observed scalar measured in the collider experiments
    from that of SM Higgs. This will also constrain the
    parameters of the model. 
    
    \item Since the masses of the DM and the vectorlike
    quarks are lighter or near TeV range, they can
    potentially be produced at the LHC. Nonobservation of
    such particles will limit the model parameter space.
\end{enumerate}
The rest of this section discusses these types of
experimental constraints in detail.

\subsection{Dark matter relic abundance}
After the $U(1)_{\rm PQ}$-symmetry breaking, the axion $a$,
being a Nambu-Goldstone, enjoys a continuous shift symmetry.
This symmetry is broken explicitly as a result of the chiral
symmetry breaking in the QCD sector, and a
temperature-dependent potential for the axion is generated
from nonperturbative QCD effects\,\cite{Gross:1980br}. But
the axion field does not start oscillating in the potential
and remains frozen at its initial value until its mass
becomes larger than the Hubble expansion rate
$H(t)=\dot{R}/R$, where $R(t)$ is the scale factor of the
Universe. After the epoch when $m_a(t)\simeq H(t)$, the
field starts oscillating coherently and the axion particles
are produced with nonrelativistic speed. They contribute
toward the CDM abundance today and their density is
approximately given by\,\cite{Abbott:1982af,Bae:2008ue},
\begin{equation}
    \Omega_ah^2 \simeq 0.18\,\theta_a^2\left(\frac{F_a}{10^{12}{\rm\ GeV}}\right)^{1.19}.
    \label{eq:axionrelic}
\end{equation}
Here $\theta_a$ is the initial misalignment angle of the
axion field relative to the minimum of the axion potential.
For simplicity, we shall assume $\theta_a\sim 1$ in the rest
of the analysis in this paper\,\cite{Sikivie:1982qv}. In
order that the axions do not overproduce DM in the Universe,
the PQ breaking scale $F_a$ has to be less than
$10^{12}{\rm\ GeV}$. In this work, we will focus on
$10^{10}\,{\rm GeV}\leq F_a \leq 10^{12}\,{\rm GeV}$.

As already noted, $\chi_1$ gains stability from the residual
$\mathbb{Z}_2$ symmetry and is a DM candidate. In the early
Universe, $\chi_{1,2}$ are in chemical equilibrium with the
thermal bath of the SM particles. As the temperature of the
Universe decreases below $\sim \mchi/20$, their rate of
interaction drops below the expansion rate and $\chi_{1,2}$
cease being in equilibrium with the SM particles. The
heavier component $\chi_2$, however, does not remain stable
as it decays to $\chi_1$, which then forms the relic
abundance $\Omega_\chi h^2$. The relic abundance is formed
after the freeze-out of $\chi_1\chi_1$ annihilations. The
annihilation can be mediated by $h$ as well as $\sigma$.
However, the $h$-mediated process dominates, since
$m_\sigma \gg m_h$. The relic abundance, being governed by
$\chi_1\chi_1\to{\rm SM\ SM}$, depends directly on $\mchi$.

\begin{figure*}
\includegraphics[width=0.4\textwidth]{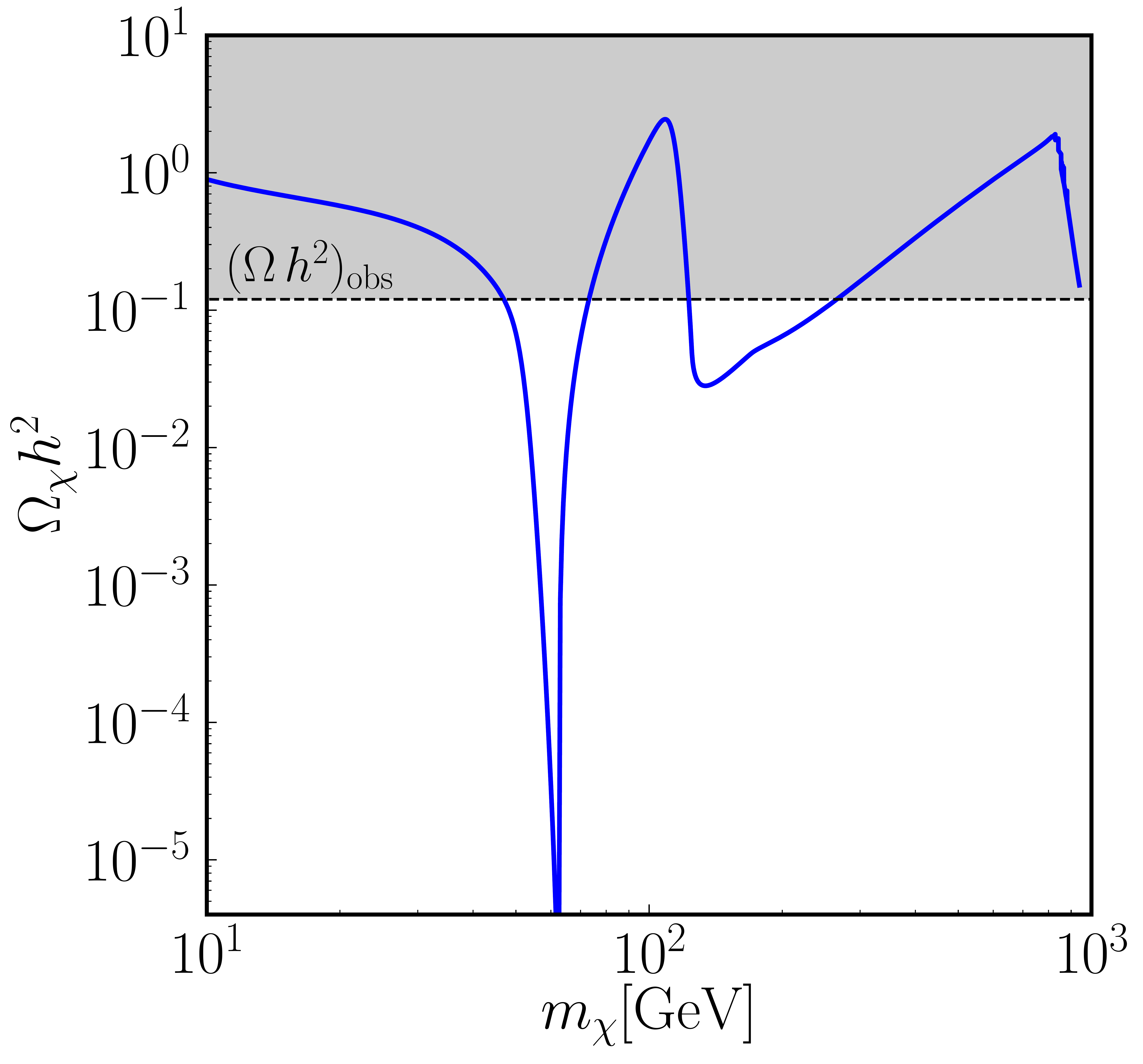}~~
\includegraphics[width=0.37\textwidth]{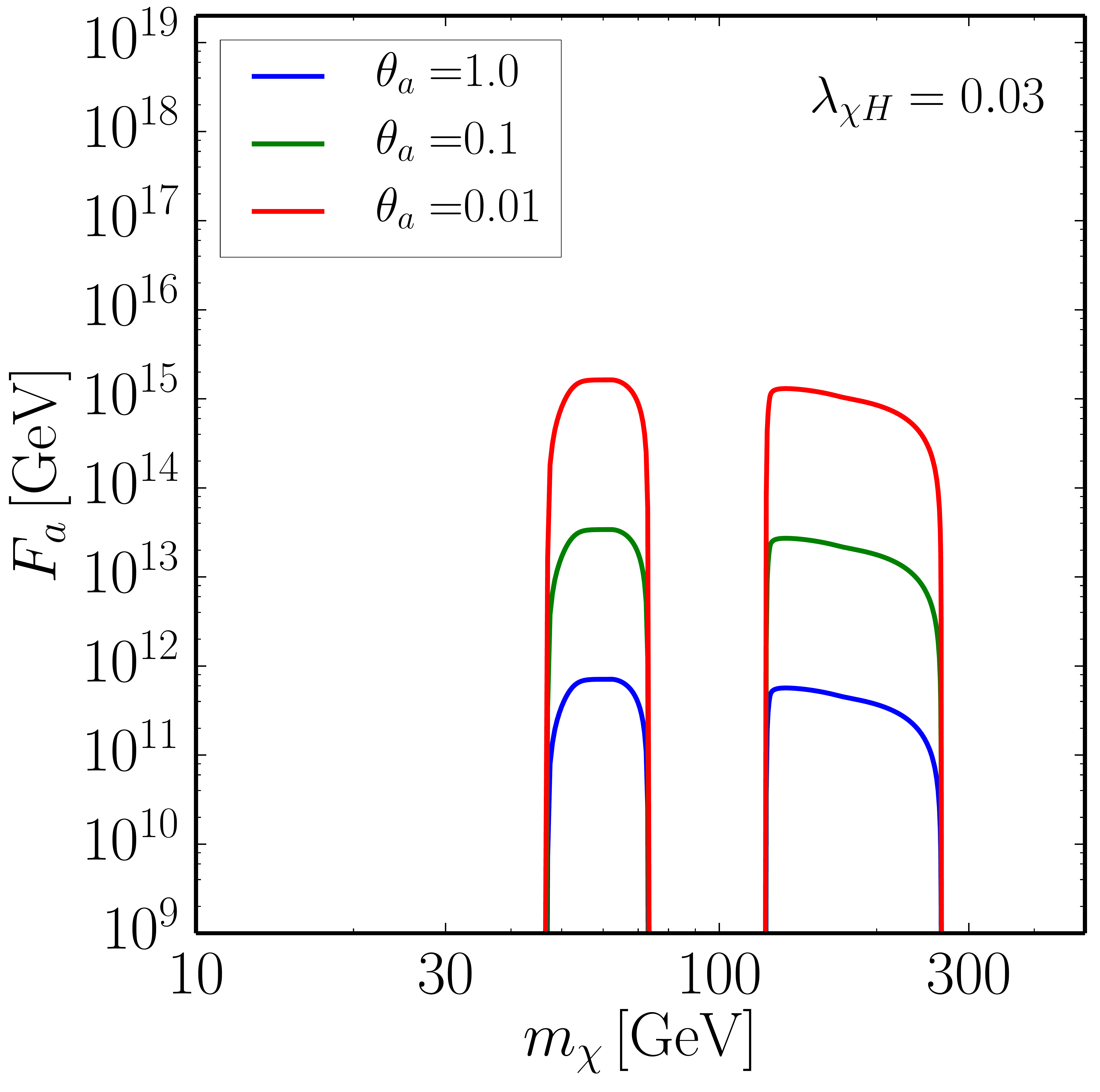}
\caption{(Left) The behavior of $\Omega_\chi h^2$ as a
function of $\mchi$. The dip at $\mchi\simeq 62.5$ GeV is
due to the s-channel resonance from $h$. The broader valley
starting from $\mchi\simeq 125$ GeV is due to opening up of
the $\chi_1 \chi_1 \to hh$ channel. The shaded region above
the $\Omega_c h^2=0.12$ line is ruled out by the Planck
experiment\,\cite{Ade:2015xua}. We allow the underabundance
regions as the axion may account for the rest of the relic
abundance. Other parameters chosen for this plot are as
follows:
$F_a=10^{10}\,{\rm GeV},\,M_Q=1\,{\rm TeV},\,f_d=0.1,\,\lambda_{\chi H}=0.03$,
and $\lambda_{\zeta H}=\lambda_{\zeta\chi}=0.1$. (Right) The
PQ scale $F_a$ needed for $\Omega_ah^2$ to satisfy the relic
constraint $\Omega_\chi h^2 + \Omega_ah^2=0.12$ for three
different values of the misalignment angle $\theta_a$.}
\label{fig:reliconly}
\end{figure*}
 
The large mass split between the two states prohibits the
possibility of coannihilation of $\chi_1$ and $\chi_2$
during DM abundance formation. As noted before, the mass
split $\Delta_\chi\sim \sqrt{F_a\epsilon_\chi}$ which, in
the region of the parameter space of our interest, is much
larger than $\mchi$. For example,
$\epsilon_\chi=1\,{\rm MeV}$ and $F_a=10^{10}\,{\rm GeV}$
imply $\Delta_\chi=1\,{\rm TeV}$. During freeze-out of
$\chi_1$, its typical kinetic energy is of order
$T_\chi \sim \mchi/20$. Therefore, by this time the number
density of $\chi_2$ particles is Boltzmann suppressed
relative to $\chi_1$,
$n_2/n_1\sim \exp{\left(-\Delta_\chi/T_\chi\right)}$ and
hence is negligible. The DM relic abundance forms only
through annihilations of $\chi_1$ into SM particles.

We show the dependence of the $\chi_1$ relic density as a
function of its mass $\mchi$ in the left panel of
Fig.\,\ref{fig:reliconly}. We used
{\tt micrOMEGAs5.0}\,\cite{Belanger:2018ccd} to numerically
compute $\Omega_\chi h^2$. The behavior for very small and
large $\mchi$ can be understood as follows. For very small
values of $\mchi(\sim 10{\rm\ GeV})$, $\chi_1$ can
annihilate only into the lighter quarks and the cross
section is suppressed by the small Yukawa couplings
resulting in overabundance of $\chi_1$. For $\mchi \gg m_t$,
the annihilation cross section is $1/\mchi^2$ suppressed.
Since the relic abundance is inversely proportional to the
annihilation cross section, we expect the region around
$\mchi\approx 100$ GeV to give the correct ballpark value of
the desired relic abundance.
 
The sharp dip at $\mchi\simeq m_h/2\simeq62.5$ GeV is due to
the s-channel resonance from the $h$ propagator. As $\mchi$
increases further from 62.5~GeV, the cross section falls
leading to a sharp increase in the relic. When the $\chi_1$
is heavier than $h$, the new annihilation channel
$\chi_1\chi_1\to hh$ opens up and dominates over all other
channels. As a result, the relic abundance decreases,
leading to the second dip. As $\chi_1$ becomes more massive,
the relic increases again because of the decrease in
annihilation cross section with the characteristic
$1/m_\chi^2$ suppression. Note that we do not consider
$\mchi>M_Q$, as the colored $Q_{L,R}$ become the lightest
dark sector particle.  

In our analysis, we take the Planck (TT, TE, EE, lowP)
measurement of the CDM energy density
$\Omega_c h^2=0.12 \pm0.0012$ represented by the horizontal
line labeled as $(\Omega h^2)_{\rm obs}$ in the left panel
of Fig.\,\ref{fig:reliconly}\,\cite{Ade:2015xua}. The
overabundance region, shown as a gray shade, is disallowed.
However, the underabundance region is allowed since the
axion abundance $\Omega_ah^2$ can account for the rest of
the relic. Therefore, the observed relic abundance
\begin{equation}
	\Omega_ch^2=\Omega_\chi h^2+\Omega_ah^2\,. \label{eqn:totalrelic}
\end{equation}
We note that $\Omega_\chi$ is virtually independent of $F_a$
due to the $v_H/F_a$ suppression in the couplings and mixing
angle. Hence, $F_a$ is fixed by \eqref{eqn:totalrelic} via
the $\Omega_a h^2$ term. 
 
In the right panel of Fig.\,\ref{fig:reliconly}, we show the
variation of $F_a$ with $m_\chi$ for three different values
of $\theta_a$, for which the axion satisfies the relic
constraint in the region, where the $\chi$ relic is
underabundant. The result for different $\theta_a$ is just a
rescaling of the value of $F_a$ according to
\eqref{eq:axionrelic}. The gap in between the allowed lines
account for the overabundance of DM due to the $\chi$.

We want to emphasize here that the interaction between
$\zeta$ and $\chi$ fields does not affect the relic
abundance of the axion and WIMP sectors. The
$\epsilon_\chi\zeta^*\chi^2 + {\rm h.c.}$ terms in
Eq.(\ref{eqn:lagrangian}) introduce interactions between
$\chi_{1,2}$ and $a$, such as $\epsilon_\chi a\chi_1\chi_2$,
$(\epsilon_\chi/F_a)a^2\left(\chi_1^2-\chi_2^2\right)$ etc.
The interaction involving $\chi_2$ is not important as its
population is already Boltzmann suppressed. The interaction
with $\chi_1$ is $(\epsilon_\chi/F_a)$ suppressed and,
therefore, not relevant for relic calculation of either
species.

\subsection{Direct detection of dark matter particles}
The DM direct detection experiments look for scattering
between the DM particles and nuclei in the detector
material. Any interaction between the DM and the SM quarks
or gluons in a given model leads to a possible signal in the
direct detection experiments. Nonobservation of such a
scattering signal in such experiments constrains the
parameters of the model. In the present case, the dominant
channel of interaction arises again through the
$h\chi_1\chi_1$ coupling, since $h$ mediates the DM and SM
quark scatterings.

\begin{figure*}
\includegraphics[width=0.41\textwidth]{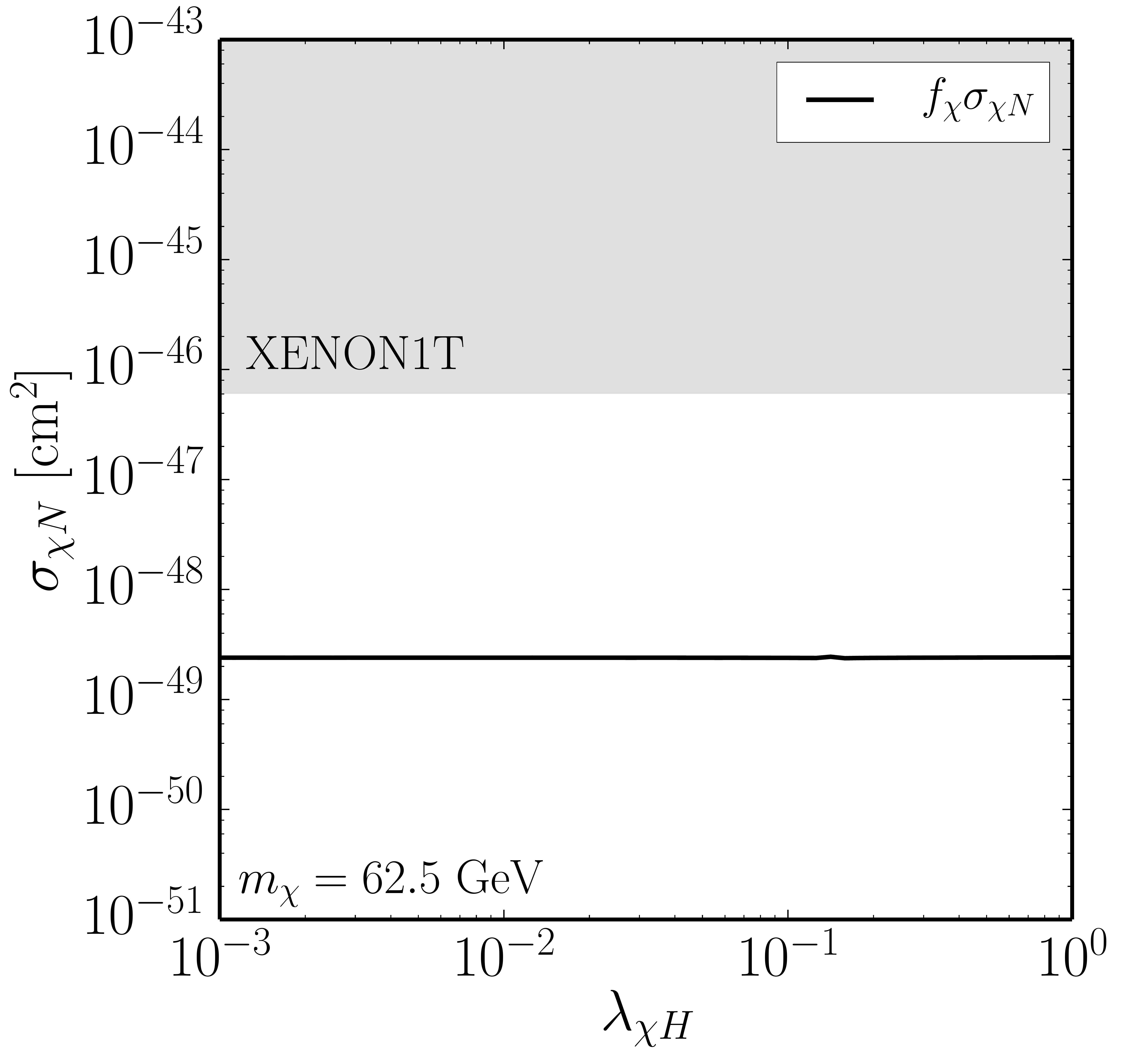}~~
\includegraphics[width=0.41\textwidth]{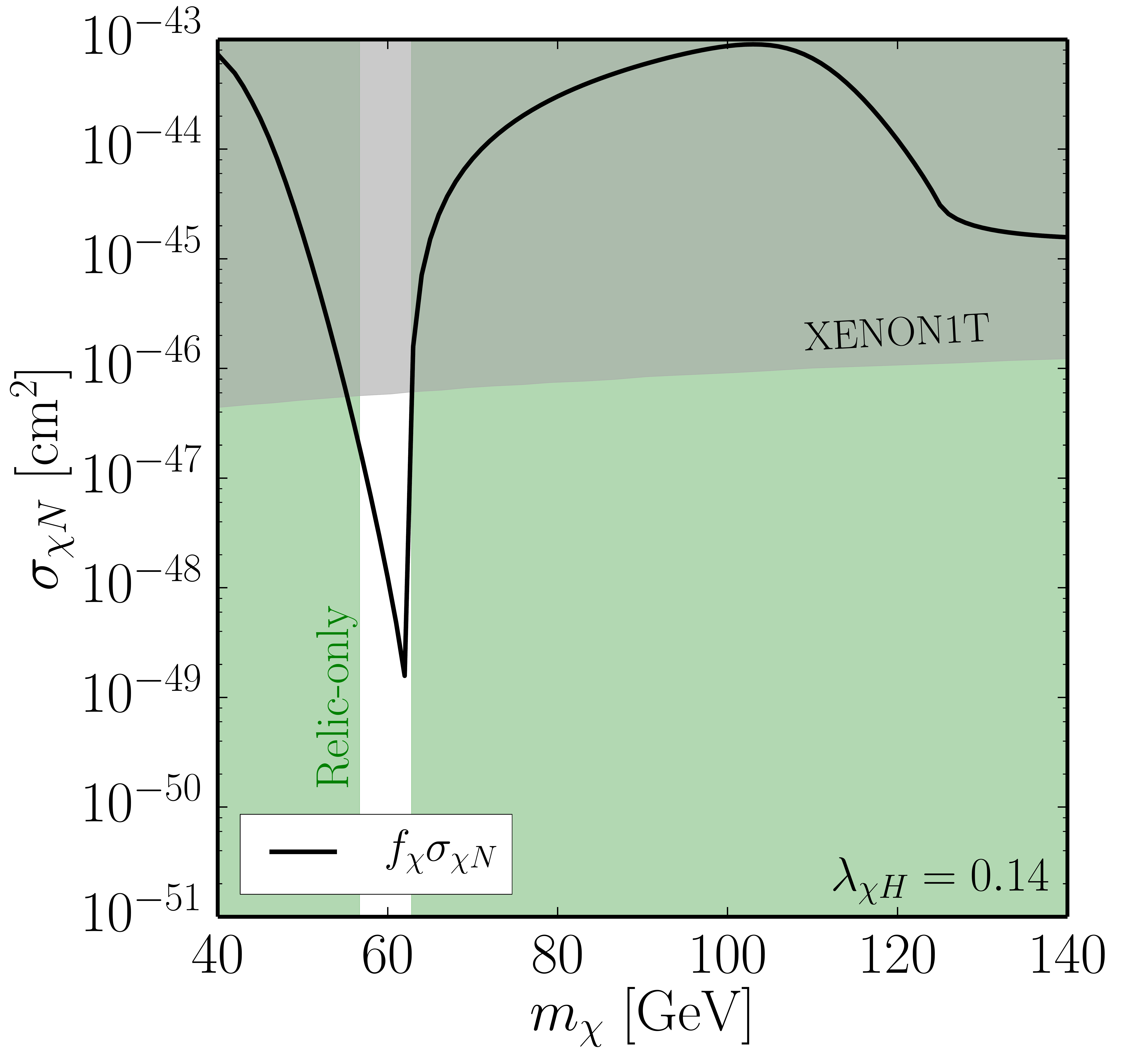}
\caption{(Left) The rescaled $\chi_1$-nucleon scattering
cross section $f_\chi\sigma_{\chi N}$ as a function of the
coupling strength $\lambda_{\chi H}$. The gray shaded region
shows the XENON1T upper bound for DM mass $\mchi=62.5$~GeV.
(Right) The rescaled $\chi_1$-nucleon scattering cross
section as a function of $\mchi$ for
$\lambda_{\chi H}=0.14$. The XENON1T experiment excludes 
the gray shaded region. The green regions are excluded by
the relic constraint itself, i.e., $f_\chi > 1$.}
\label{fig:ddlchih}
\end{figure*}

The DM-nucleon scattering cross section $\sigma_{\chi N}$ is
constant for very small $\lambda_{\chi H}$ because the
coupling becomes independent of $\lambda_{\chi H}$. For very
large $\lambda_{\chi H}$, the cross section increases as
$\sim \lambda_{\chi H}^2$, as expected. In between, a dip
occurs because of the cancellation of two terms appearing in
the vertex factor of $h\chi_1\chi_1$ coupling [see
\eqref{eqn:approxhChiChi1}]. Note that this cancellation is
entirely due to kinematics. There is neither any dynamical
symmetry imposed to keep $m_\chi$ around $m_h/2$ nor any
fine-tuning required. Since the enhancement of the cross
sections is due to kinematics, the enhancement will remain
stable under radiative corrections. Note that in this model,
$\chi_1$ forms only a fraction $f_\chi$ of the total dark
matter
abundance\,\cite{Cao:2007fy,Bhattacharya:2013hva,Ilnicka:2015jba,Bhattacharya:2016ysw,Ahmed:2017dbb,Betancur:2018xtj,Borah:2019aeq}:
\begin{equation}\label{eq:fchi}
f_\chi= \frac{\Omega_\chi}{\Omega_c}\,.
\end{equation}
Therefore the DM-nucleon cross section needs to be
\emph{rescaled} by $f_\chi$ before comparing it with the
experimental results.

Note that in the literature, another convention of rescaling
the DM-nucleon scattering cross section given by
$f_\chi = \mathrm{Min}[1, \Omega_\chi/\Omega_c]$ exists,
which saturates $f_\chi$ to unity in the overabundant
region\,\cite{PhysRevD.34.2206, Bottino:1996eu}. However, in
the context of our model, we use the prescription given in
\eqref{eq:fchi}. This is well justified, since there exists
a concrete prediction for calculating the DM relic in this
model. Particularly, when there is a global overdensity, we
do not assume that any other unknown mechanism can account
for the relic density locally. While the direct and indirect
detection constraints would depend on the choice of
$f_\chi$, considering them with the relic bounds does not
yield any additional allowed regions. Hence, our definition
does not affect the final results.

Presently, the most stringent bound on the DM-nucleon cross
section is given by the XENON1T$\times$1 yr
data\,\cite{Aprile:2018dbl}. It is most sensitive to the DM
mass in the range  $10\,{\rm GeV}-1\,{\rm TeV}$ and the
strongest upper bound quoted is
$\sigma_{\chi N}\simeq 10^{-46}{\rm\ cm^2}$. The rescaled
cross section $f_\chi\sigma_{\chi N}$ as a function of
$\lambda_{\chi H}$ is shown in the left panel of
Fig.~\ref{fig:ddlchih}. The cross section has a dip at
$\lambda_{\chi H}=0.14$ and increases as
$\sim\lambda_{\chi H}^2$ for larger values as explained
before. However, $f_\chi$ on the other hand has a peak at
the same parameter point because of inefficient relic
annihilation due to vanishing of the $h\chi_1\chi_1$
coupling. Additionally, it has an inverse relation with
$\lambda_{\chi H}$ for larger values:
$f_\chi\sim \sigma_\mathrm{ann}^{-1} \sim \lambda_{\chi H}^{-2}$.
Therefore, together the rescaled cross section
$f\chi\sigma_{\chi N}$ does not have any features as shown
in the left panel in Fig.~\ref{fig:ddlchih}.
 
We will show later that due to the stringent constraint, the
only experimentally allowed region of DM mass turns out to
be around $\mchi\simeq 62.5$ GeV. This is shown in the right
panel of Fig.~\ref{fig:ddlchih}, where we plot
$f_\chi\sigchiN$ as a function of $\mchi$ for
$\lambda_{\chi H}=0.14$. The gray region shows the XENON1T
constraint. In passing, we comment that the region around
$m_\chi\sim m_h$ is allowed by the relic constraint only if
$\lambda_{\chi H} \lesssim 0.079$ or
$\lambda_{\chi H} \gtrsim 0.25$. However, these regions are
excluded by the XENON1T bound.

All the above bounds apply for $\chi_1$ as the DM candidate.
However, direct detection experiments for axion need to
follow a different search strategy because of its ultralow
mass. There have been a few experimental efforts to look for
axionic dark matter. For example, the ADMX
experiment\,\cite{2010PhRvL.104d1301A} uses a rf cavity to
look for its interaction with the electromagnetic field. In
the KSVZ model, the interaction strength between an axion
and two photons is given
by\,\cite{Kim:1979if,Shifman:1979if}
\begin{equation}
g_{a\gamma}=-1.92\frac{\alpha}{2\pi F_a}\,,
\end{equation}
where $\alpha$ is the fine structure constant. Presently,
ADMX rules out a narrow region of the parameter space above
$g_{a\gamma}\simeq 10^{-15}{\rm\ GeV^{-1}}$
($F_a\simeq 10^{12}$~GeV) around $m_a\simeq 2\ \mu$eV. For
a higher mass axion, the bound is even weaker. Another
proposed experiment is CASPEr-Electric which will probe
$F_a\gtrsim 10^{12}$ GeV for lighter
axions\,\cite{Budker:2013hfa}. Moreover, we should remember
that these bounds assume that 100\% CDM abundance is given
by axion which is not be true in our model. These bounds are
weaker than the upper limit on $F_a$ from the dark matter
relic abundance, even after adjusting for the correct factor
to cancel out the assumption and, hence, do not require
special attention.

\subsection{Dark matter annihilation signal}
Various astrophysical observations hint that the present day
Universe consists of galaxies sitting inside halolike
structures formed by gravitational clustering of DM
particles\,\cite{Clowe:2006eq}. At the center of these
halos, the DM density is high enough to scatter with each
other and annihilate into SM particles. These final state
particles would further decay and give rise to gamma-ray
signals from various astrophysical objects, such as dwarf
galaxies, the Milky Way center etc. We focus on bounds
arising from gamma-ray signals due to such annihilations of
DM particles.

\begin{figure}[h]
\includegraphics[width=0.42\textwidth]{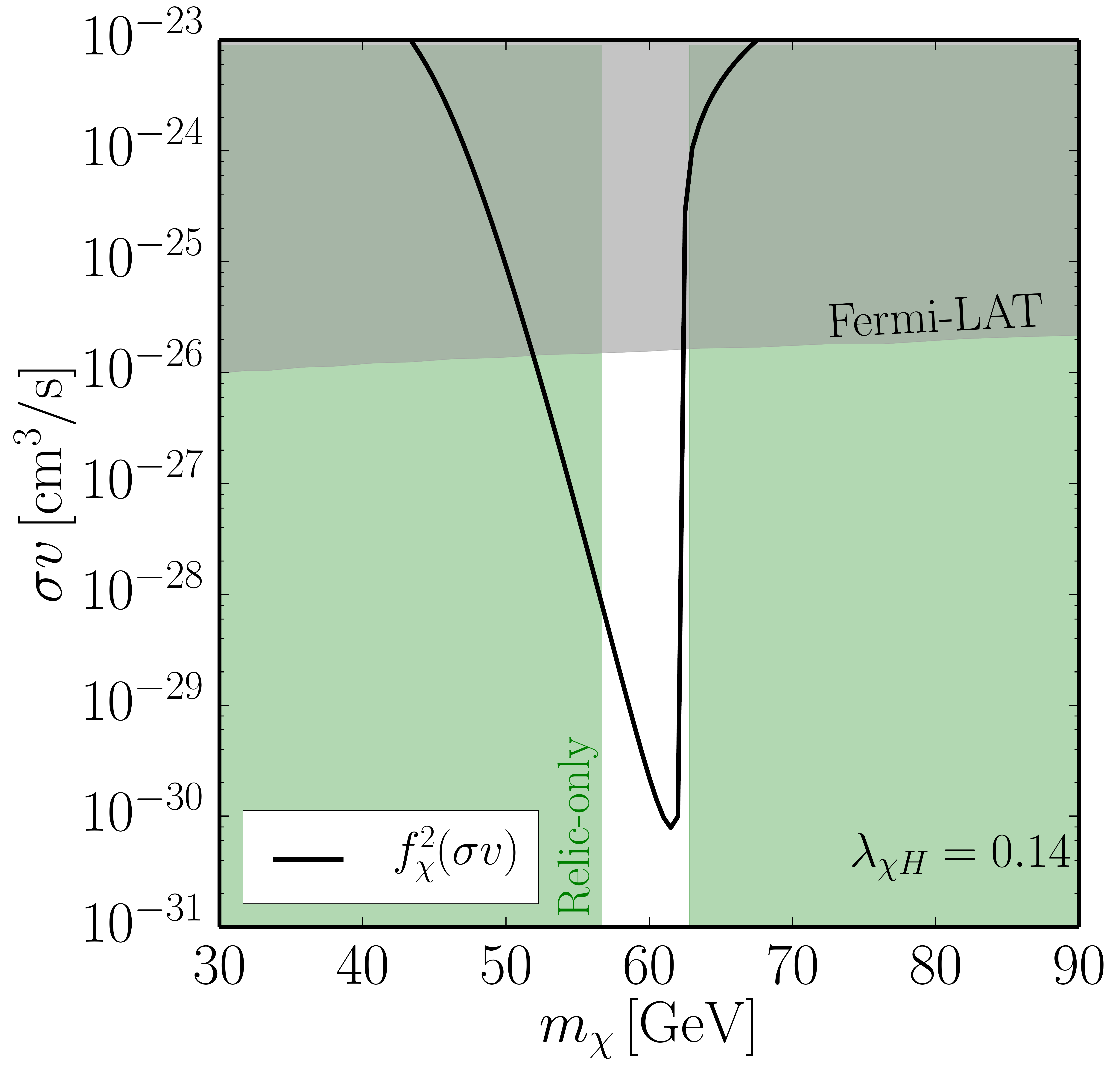}
\caption{The rescaled annihilation rate $f_\chi^2(\sigma v)$
of $\chi_1\chi_1$ into $b\bar b$ in this model as a function
of the mass of $\chi_1$  for $\lambda_{\chi H}=0.14$. The
sharp dip is due to the s-channel resonance from the SM
Higgs. The most stringent upper bound on this cross section
is provided by the dwarf galaxy observation of the Fermi-LAT
satellite data\,\cite{Fermi-LAT:2016uux}. The gray shaded
region is ruled out by the Fermi-LAT constraint. The green
regions are excluded by the relic constraint itself, i.e.,
$f_\chi > 1$.}\label{fig:idmchi}
\end{figure}

We pay more attention to the DM mass around
$\mchi\simeq m_h/2= 62.5$ GeV which is still allowed by the
direct detection experiment data. The total annihilation is
dominated by the $b\bar{b}$ channel ($\sim 90\%$). The
rescaled annihilation rate $f_\chi^2(\sigma v)$ is shown in
Fig.\,\ref{fig:idmchi} as a solid black line. Note that here
also the annihilation cross section is enhanced due to the
s-channel resonance from the SM Higgs propagator. However,
after rescaling with $f_\chi^2$, which is decreased at
around $m_\chi\approx 62.5$ GeV, the annihilation rate
$\sigma v$ shows a dip followed by a sharp increase around
that point.  The dependence on $\lambda_{\chi H}$ comes
through the $g_{h\chi_1\chi_1}$ coupling.

There have been many experiments which have looked for DM
annihilation signals from various astrophysical
objects\,\cite{Fermi-LAT:2016uux,Abdallah:2016ygi,Ahnen:2017pqx,Aartsen:2017ulx}.
At present, the most stringent upper bounds on the thermally
averaged DM annihilation cross section
$\langle\sigma v\rangle$ is given by the DES-Fermi-LAT joint
gamma-ray search from the satellite galaxies of the Milky Way\,\cite{Fermi-LAT:2016uux}. It is derived from 6 yr
observation of 45 such objects by the LAT. They have
relatively less amount of visible baryonic matter and the DM
population is expected to dominate their matter density. In
Fig.~\ref{fig:idmchi}, we show this upper bound on the
annihilation cross section due to Fermi-LAT as the gray
shaded region.  Note that the indirect detection bounds rule
out most part of our parameter space, except a region around
$\mchi\simeq m_h/2$. In passing, we also note that the DM
mass needed for the resonantly enhanced annihilation signal
in the $b\bar{b}$ channel matches the result of the
\textit{Galactic center excess} analysis done in
Ref.\,\cite{Huang:2013pda} within $1\sigma$ C.L. (also
see\,\cite{Calore:2014xka}).

\subsection{New physics searches at the LHC}
In this subsection, we will focus on various signatures of
the model at the LHC. The model has an extended scalar
sector: apart from the SM Higgs boson $h_0$, there exists a
scalar DM candidate $\chi_1$ and its heavier counterpart
$\chi_2$, and another scalar field $\sigma_0$, which is the
radial component of $\zeta$. As discussed earlier, $h_0$ and
$\sigma_0$ mix with each other giving rise to physical
states $h$ and $\sigma$. The mixing between $\sigma_0$ and
$h_0$ changes the properties of $h$ from that of the SM
Higgs via its coupling to SM particles as well as to the new
states present in this model. Since various properties of
the observed scalar particle at the LHC resemble that of the
SM Higgs boson, we expect some constraints on the parameter
space of the model from the measurement of the properties of
the observed 125~GeV scalar. 

One of the measurements that provides relevant information
about the properties of the observed 125 GeV scalar is its
signal strength. If the scalar decays to
$X \in \{\ell^\pm, q, g, Z, W$\} and its conjugate $\bar X$,
its signal strength is defined as
\begin{equation}
    \mu_{X\bar X} = \frac{\sigma_\text{exp}(pp \rightarrow h)\times\text{BR}_\text{exp}(h\rightarrow X\bar X)}{\sigma_\text{SM}(pp \rightarrow h)\times\text{BR}_\text{SM}(h\rightarrow X\bar X)}\,, \label{eqn:SS-def}
\end{equation}
where $\sigma_\text{exp}$ stands for the experimentally
observed cross section of the process $pp \rightarrow h$ and
$\text{BR}_\text{exp}$ is the experimentally observed
branching ratio of the process $h\rightarrow X\bar X$.
Similarly, $\sigma_\text{SM}$ and $\text{BR}_\text{SM}$ in
\eqref{eqn:SS-def} stand for the corresponding values
predicted in the SM. We compare observed $\mu_{X\bar X}$
with the theoretically calculated $\mu_{X\bar X}$ from the
model in different decay channels.

Due to the mixing, the physical scalar $h$ will have a
$\cos\theta$ component in all the couplings with the SM. An
additional decay mode of $h$ to $\chi_1\chi_1$ is possible
if $\mchi < m_h/2$. If the partial decay width of the new
decay modes of $h$ is $\Gamma^\text{new}$, the signal
strength of $h$ decaying to any SM particle pairs $X\bar X$
can be written as 
\begin{equation}
    \mu_{X\bar X} = \dfrac{\cos^2\theta}{1 + \dfrac{\Gamma^\text{new}}{\cos^2\theta\ \Gamma^\text{tot}_\text{SM}}}\,, \label{eqn:SS-model}
\end{equation}
where $\Gamma^\text{tot}_\text{SM}$ is the total decay width
of SM Higgs boson.

\begin{ruledtabular}
\begin{table}[h]
\caption{\small Measured values of the signal strengths of
the 125~GeV observed scalar. The superscripts represent the
production modes and the subscripts indicate the decay modes
of the observed scalar $h$. The measurements are done by
ATLAS and CMS at the LHC with $\sim 36$~fb$^{-1}$ luminosity
at $\sqrt{s}=13$~TeV.}\label{tab:higgs-properties}
\begin{tabular}{lcc}
 && \\[-10pt]
&    ATLAS                 &    CMS    \\[1pt]
\hline && \\ [-10pt]
$\mu^\text{(ggF)}_{W W^*}$ & $1.21^{+0.22}_{-0.21} $\,\cite{ATLAS:2018gcr}  &   $1.38^{+0.21}_{-0.24}$\,\cite{CMS:2018xuk} \\[6pt]
$\mu^\text{(ggF)}_{ZZ^*}$ & $1.11^{+0.23}_{-0.21}$\,\cite{Aaboud:2017vzb}     &    $1.20^{+0.22}_{-0.21}$\,\cite{Sirunyan:2017exp}      \\[6pt]
$\mu_{\gamma\gamma}^\text{(ggF+VH+VBF+ttH)}$ & $0.99^{+0.15}_{-0.14}$\,\cite{Aaboud:2018xdt}     &    $1.18^{+0.17}_{-0.14}$\,\cite{CMS:2018lkl}      \\[6pt]
$\mu_{b\bar b}^\text{(VH)}$ & $1.20^{+0.42}_{-0.36}$\,\cite{Aaboud:2017xsd}     &   $1.06^{+0.31}_{-0.29}$\,\cite{Sirunyan:2017elk}    \\[6pt]
$\mu_{\tau \tau}^\text{(ggF+VH+VBF)}$ & $1.43^{+0.43}_{-037}$\,\cite{Aad:2015vsa} &   $1.09^{+0.27}_{-0.26}$\,\cite{Sirunyan:2017khh}    \\[4pt]
\end{tabular}
\end{table}
\end{ruledtabular}

\begin{figure*}
\includegraphics[width=0.4\textwidth]{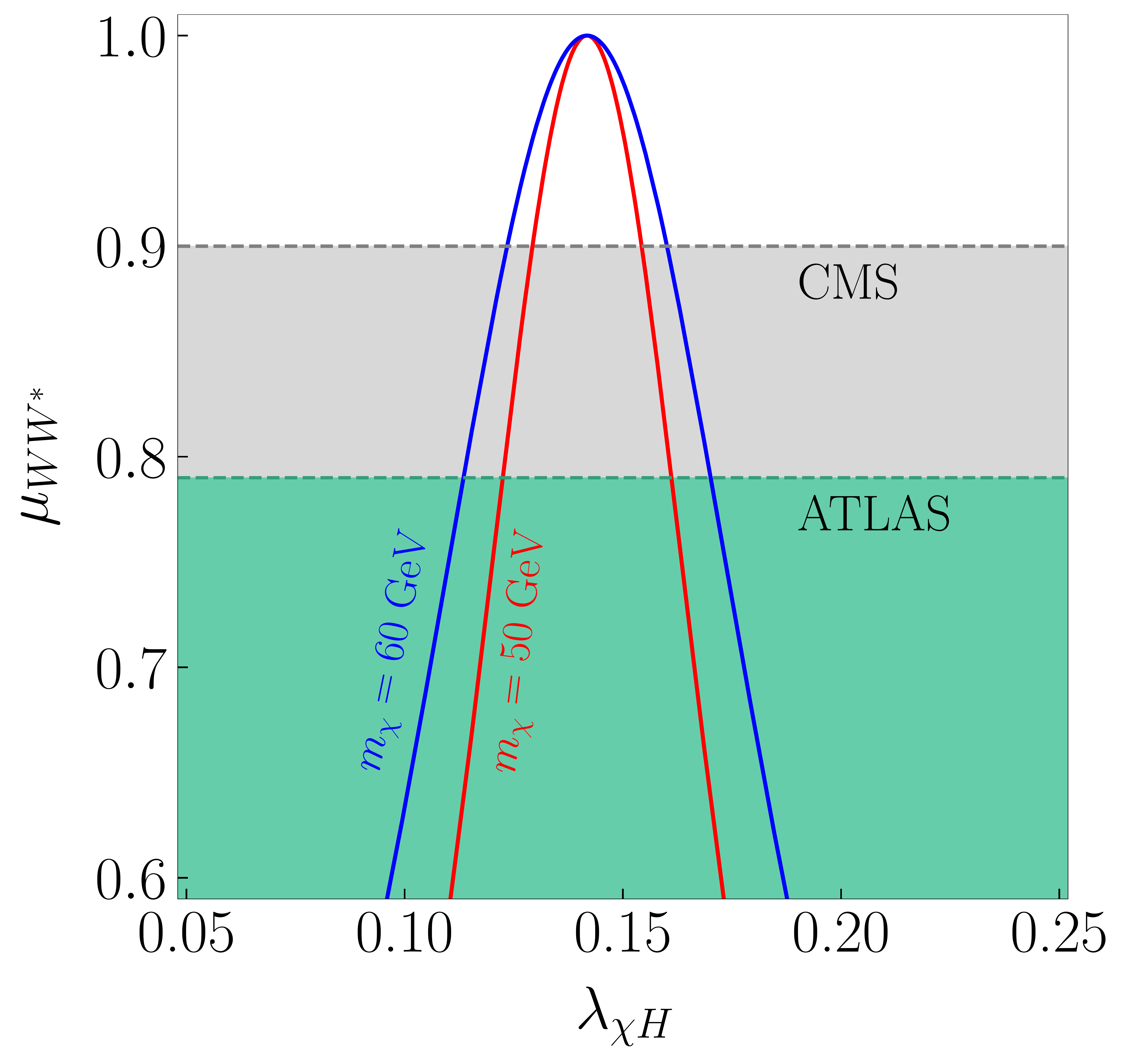}~
\includegraphics[width=0.4\textwidth]{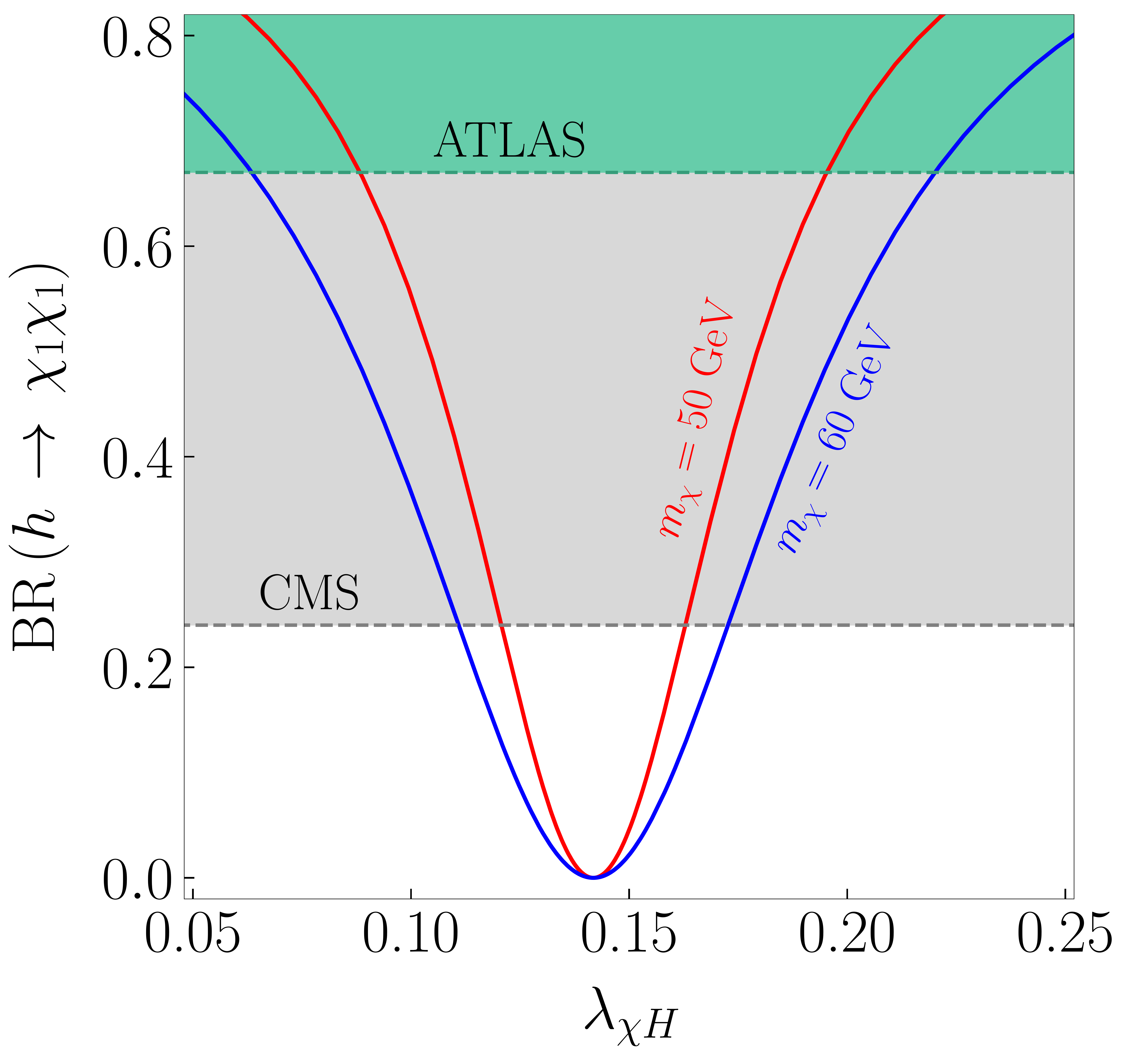}
\caption{Bounds arising from (left) the Higgs signal
strength in the $WW^*$ channel and (right) the invisible
decay of the Higgs. The gray (green) shaded regions in both
the plots are excluded by CMS (ATLAS) measurement at
95\% C.L. The allowed regions are shown in white.}
\label{fig:SS-inv}
\end{figure*}

In Table~\ref{tab:higgs-properties}, we tabulate the recent
measurements of signal strength of the observed scalar $h$
by both ATLAS and CMS Collaborations at 13 TeV with
$\sim 36$ pb$^{-1}$ integrated luminosity in different decay
channels of $h$. The superscripts in the $\mu_{X\bar X}$
represent the production mode of the scalar $h$. For our
analysis, we constrain the parameter space by imposing the
value to be at 95\% C.L. of the measured values, i.e., with
$\pm 2\sigma$ around the measured central value. Since, in
the model, $\mu_{X \bar X}$ is always below unity, it is the
lower bound at 95\% C.L. which will actually put constraints
on the parameters.

In the left panel of Fig.~\ref{fig:SS-inv}, we show the
variation of the signal strength of $h$ in the $WW^*$
channel as a function of $\lambda_{\chi H}$ for two
different masses of $\chi_1$. As expected from
\eqref{eqn:SS-model}, the variation is a Lorentzian, with a
narrow width governed by $\Gamma^\text{tot}_\text{SM}$ and
$\mchi$. Since the coupling for $h$ to $\chi_1\chi_1$, as
given in \eqref{eqn:approxhChiChi1}, vanishes at
$\lambda_{\chi H} = 2\lambda_{\zeta\chi}\left(\lambda_H - \lambda_H^{SM}\right)/\lambda_{\zeta H}$ ($\approx 0.14$
for the chosen benchmark point), the decay mode for the $h$
vanishes at that point, and hence the $\mu_{X\bar X}$
becomes 1 around that point. The gray (green) shaded region
shows the area disallowed at 95\% C.L. by the measurements
by CMS (ATLAS) as indicated in the plot, and the allowed
region is shown in white. Although the measurements for
different decay channels of $h$ are listed in
Table~\ref{tab:higgs-properties} for completeness, we only
plotted $\mu^\text{(ggF)}_{W W^*}$, which gives the
strongest bounds from the signal strength measurement.

We also study the bounds from the invisible decay of $h$
which arises from the decay channel
$h\rightarrow\chi_1\chi_1$ for $\mchi < m_h/2$ in this
model. The BR of the decay can be written as
\begin{equation}
\text{BR}(h \to \chi_1 \chi_1) = \frac{1}{1 + \cos^2\theta\,\dfrac{\Gamma^\text{tot}_\text{SM}}{\Gamma^\text{new}}}\,.
\end{equation}
The dependence of BR$(h\to\chi_1\chi_1)$ with the parameter
$\lambda_{\chi H}$ is plotted in the right panel of
Fig.~\ref{fig:SS-inv} for two different masses of $\chi_1$.
As in the case with the signal strength, the
BR$(h\to\chi_1\chi_1)$ vanishes at the point where the
coupling of $h$ to $\chi_1 \chi_1$, given by
$g_{h\chi_1\chi_1}$ [see \eqref{eqn:approxhChiChi1}], goes
to zero. This feature is evident from the plot in the right
panel of Fig.~\ref{fig:SS-inv}. Away from this point, the BR
increases in both sides, tending to unity for a high value
of $g_{h\chi_1\chi_1}$, which indicates that
$\Gamma^\text{new}$ is the dominant decay mode, and all
other modes are suppressed.

\begin{ruledtabular}
\begin{table}[h]
\caption{\small Observed upper limit on the branching ratio
of invisible decay of the scalar $h$.}\label{tab:inv-decay}
\begin{tabular}{ccc}
&    ATLAS                 &    CMS    \\
\hline	& & \\[-2.5mm]
BR$\left( h \rightarrow \text{inv}\right)$ & 0.67\,\cite{Aaboud:2017bja}     &    0.24\,\cite{CMS:2018awd}      \\
\end{tabular}
\end{table}
\end{ruledtabular}
 
Nonobservation of this decay mode of the observed $125$ GeV
scalar at the LHC, therefore, places an upper limit on the
invisible decays of $h$. These upper limits are tabulated in
Table~\ref{tab:inv-decay}. In the right panel of
Fig.~\ref{fig:SS-inv}, the gray (green) shaded region is the
area disallowed at 95\% C.L. by CMS\,\cite{CMS:2018awd}
(ATLAS\,\cite{Aaboud:2017bja}) measurements on the invisible
decays of $125$~GeV scalar. It is therefore clear that only
a small range of $\lambda_{\chi H}$, for which the BR curves
fall within the white region, is allowed by current
measurements. 

At this point, it is worth mentioning that the trilinear
coupling of $h$ is also modified due to the mixing with
$\sigma_0$, which will change the di-Higgs production rate.
Measurements for the trilinear coupling of $h$ as well as
di-Higgs production have been carried out by both
ATLAS\,\cite{ATLAS:2018otd} and CMS\,\cite{Sirunyan:2018iwt}
in the di-Higgs channel. However, the upper bounds are well
above the SM prediction due to lack of signal in the
di-Higgs channel. Hence, much of the parameter space,
especially the region of interest, of the model is not
constrained by the measurement of trilinear coupling of $h$.

\begin{figure}[h]
\includegraphics[width=0.4\textwidth]{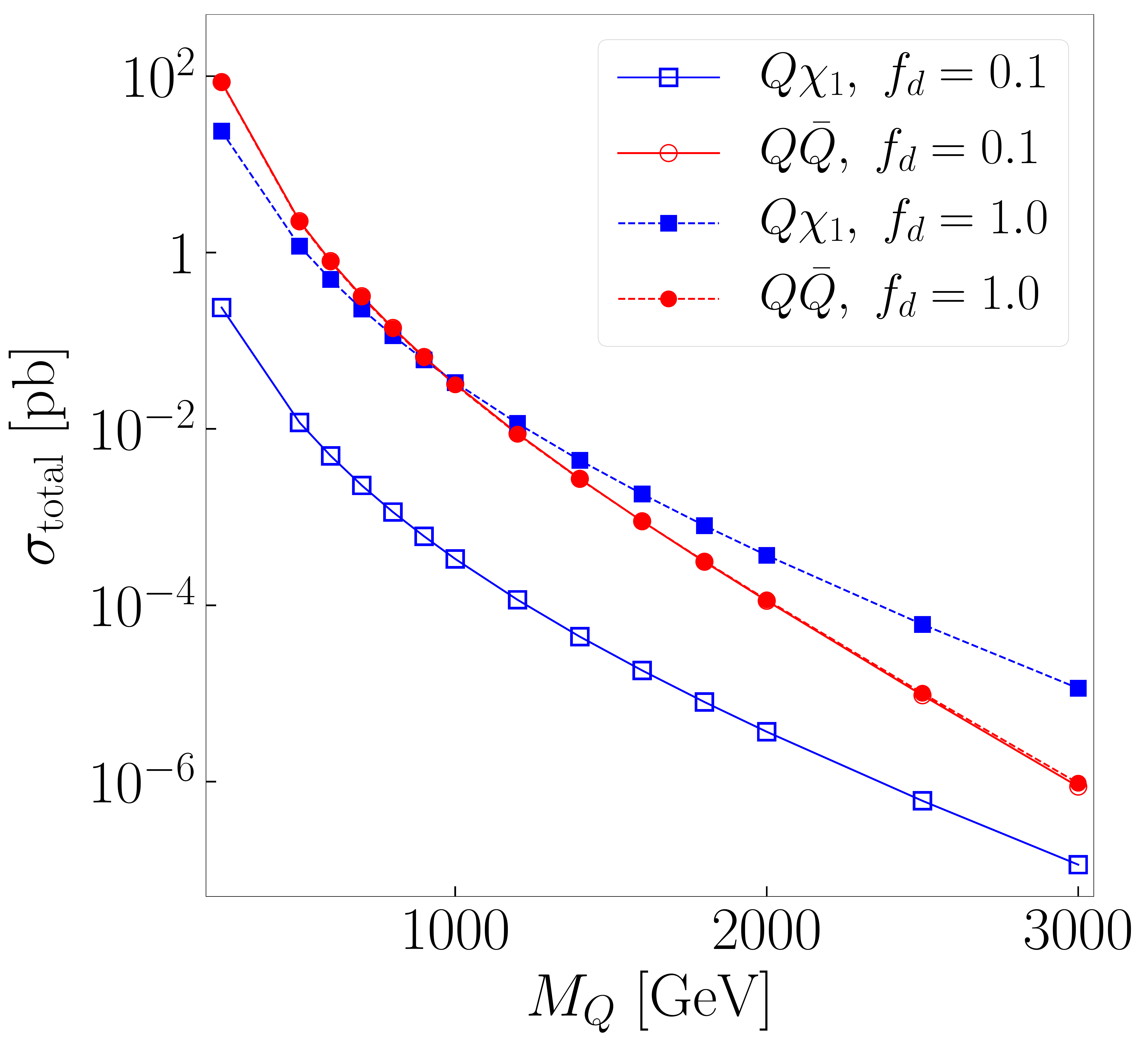}
\caption{Variation of total production cross section for
$Q\bar Q$ (in red) and for $Q\chi_1$ and $\bar Q̄\chi_1$ (in
blue) in dijet+MET and in monojet+MET channels,
respectively, at the LHC at $\sqrt{s}=13$ TeV as a function
of $M_Q$ for $m_\chi = 60$~GeV.}\label{fig:VLQ}
\end{figure}

\begin{figure*}
\includegraphics[width=0.4\textwidth]{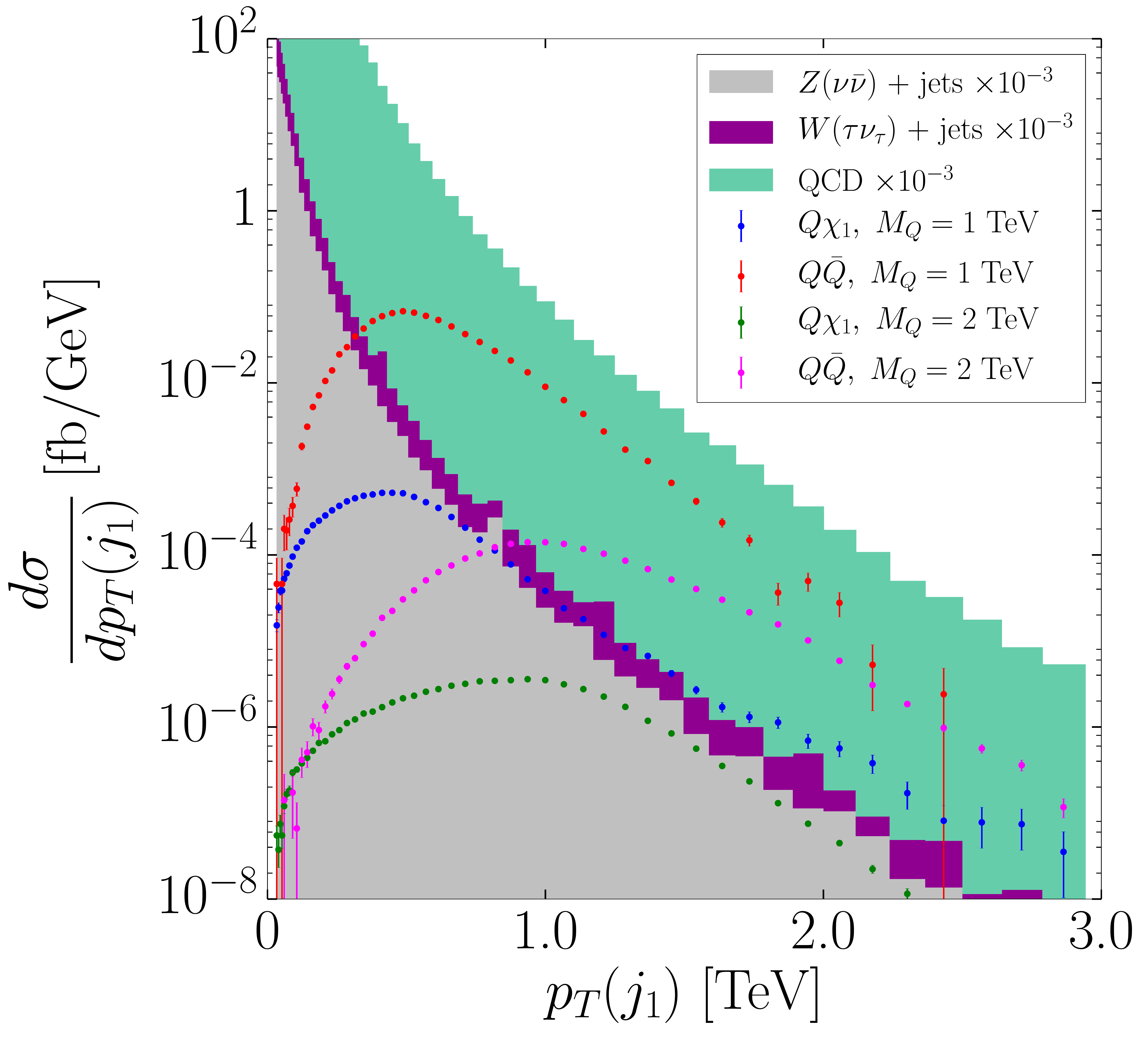}
\includegraphics[width=0.4\textwidth]{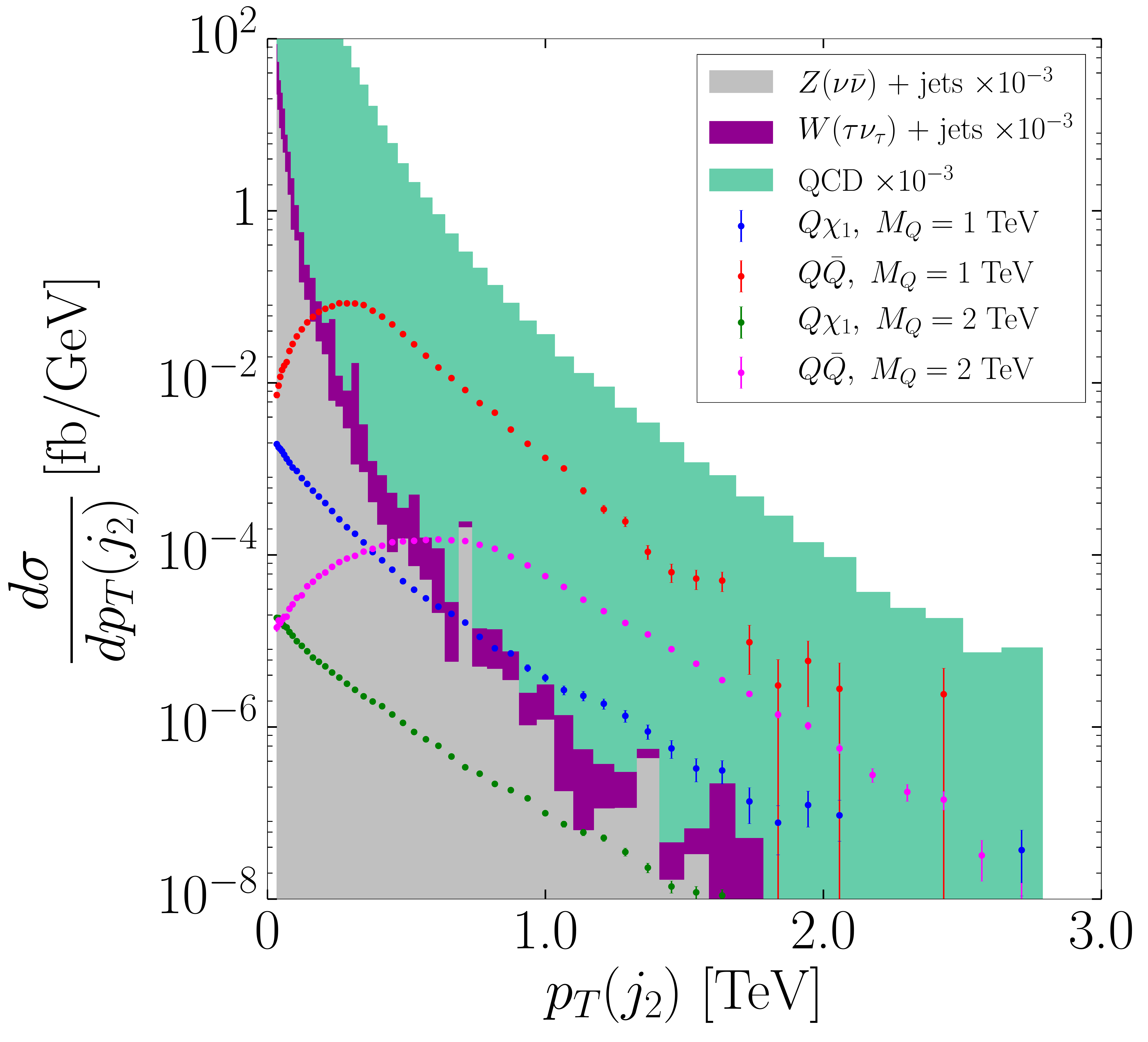}\\
\includegraphics[width=0.4\textwidth]{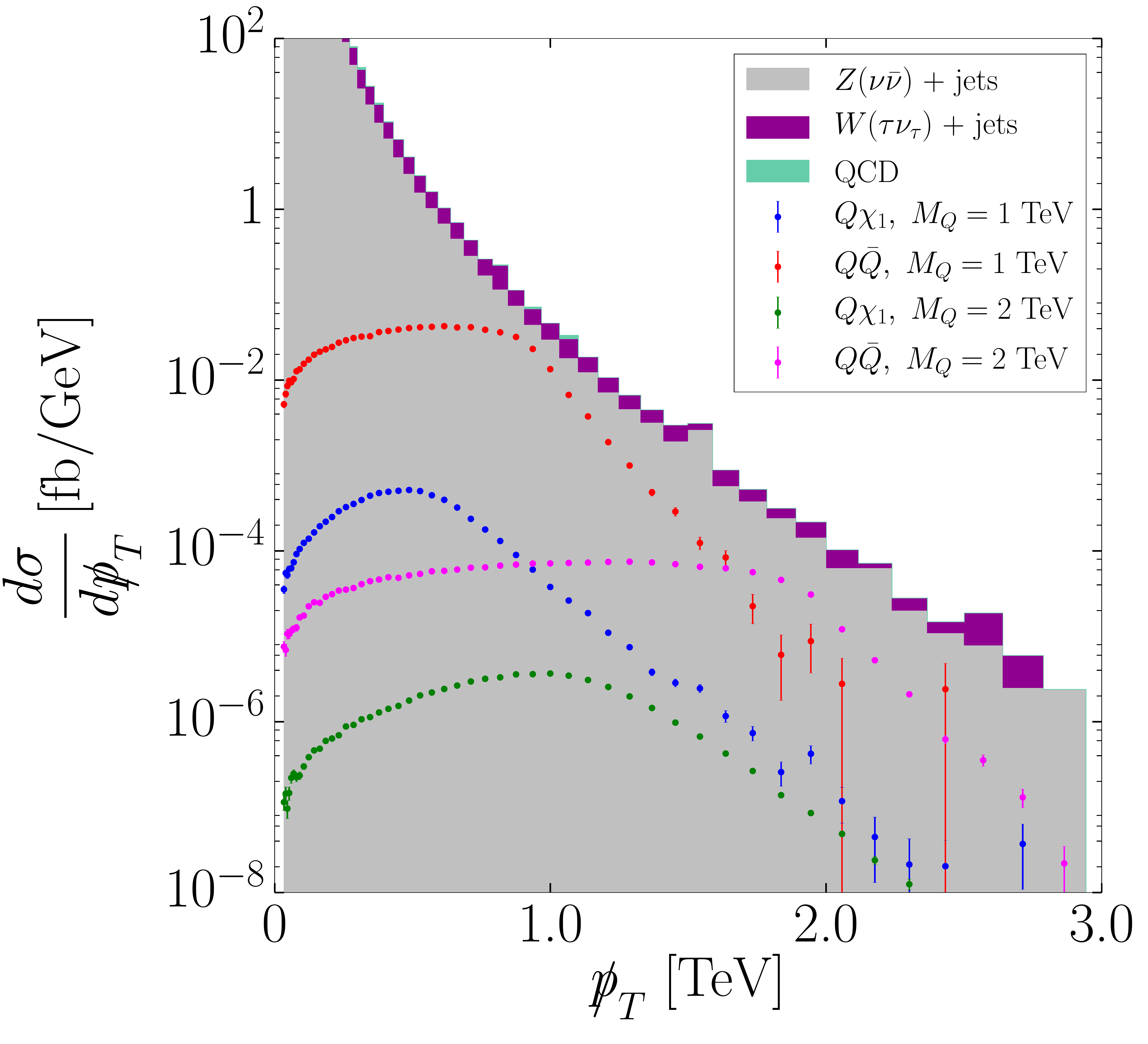}
\includegraphics[width=0.4\textwidth]{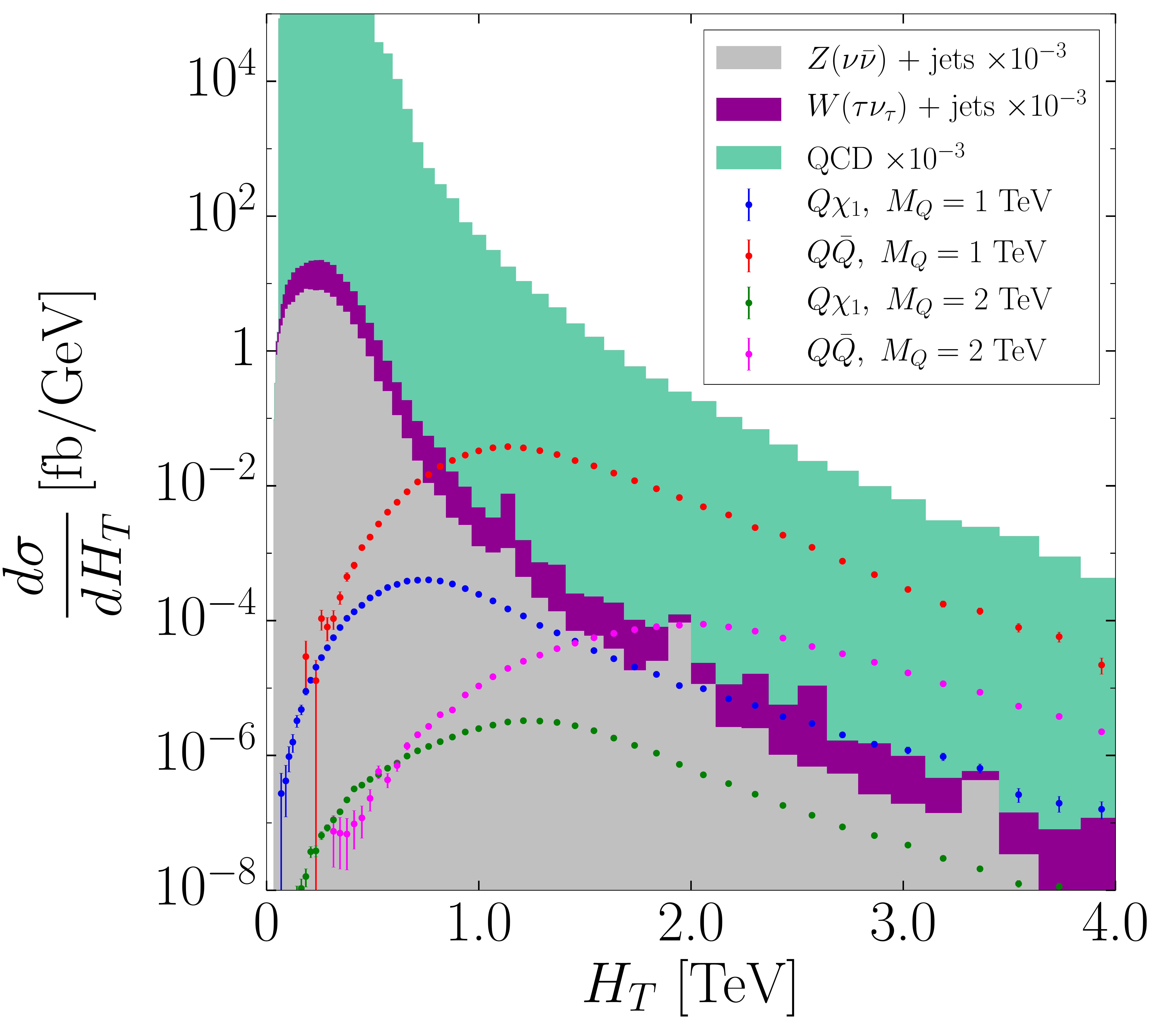}
\caption{Differential distributions of signal and background
events for dijet+MET and monojet+MET final states.
Distributions are for (top left) $p_T$ of the leading jet,
(top right) $p_T$ of the second jet, (bottom left) missing
transverse energy $\slashed p_T$, and (bottom right) the
scalar sum of $p_T$ of all the jets
$H_T =\!\!\underset{j\,\in\,\text{jets}}{\sum} |\vec p_{T_j}|$
for different values of $M_Q$.}\label{fig:dist}
\end{figure*}

The model also predicts new particles at around GeV--TeV
range, which can potentially be observed in a TeV collider.
One such particle is the DM candidate $\chi_1$, which is
weakly interacting and does not decay within the detector.
If it is produced in the collider, it will not be detected
and will contribute to the missing momentum in an event. The
other particles, within the observable range of TeV
collider, are the vectorlike quarks $Q_L$ and $Q_R$. Since
these quarks are colored, they can be produced in a hadron
collider and subsequently decay to a down-type quark and a
$\chi_1$. The presence of $\chi_1$ will again contribute to
the missing energy in the detector. The lack of agreement of
such signals with those predicted at the TeV colliders will
also put bounds on the parameter space of the model in
consideration. 

The contribution to the oblique parameters $(S,\ T$ and $U$)
matters if the vectorlike quarks mix with our SM quarks. In
that scenario, it contributes to the parameters strongly,
depending on the mixing angle\,\cite{Dawson:2012di}. In our
case, we do not directly have a SM top and $Q_{L,R}$ mixing
because of the PQ symmetry. The mixing happens only through
the $\zeta$-Higgs mixing, which induces a $h \bar{Q}_L Q_R$
term. This mixing is suppressed by $v_H/F_a$, which is super
tiny. As a result, the contributions to the $S,\ T$ and $U$
parameters are not important.

Now, we turn to the discussion of direct production of the
new particles at the LHC. The new particles, being charged
under a PQ symmetry, should be produced in pairs. There are
three different pairs of new particles that can be directly
produced: $Q\bar Q$,  $Q\chi_1$, and $\bar Q \chi_1$. Hence,
these processes will contribute to the following final
states:  dijet $(2j) + $MET  in case of $Q\bar Q$ production
and monojet $(j)+$MET final state in case of $Q\chi_1$ and
$\bar Q \chi_1 $ production, where MET stands for missing
transverse energy.  In the rest of this section, we will
discuss the constraints on the parameter space in view of
the observation of the above-mentioned final states at the
collider.

Since the $Q$s are colored, the cross section for the
production of $Q \bar Q$ will be similar to that of the SM
quarks and will be suppressed for higher masses.
Figure~\ref{fig:VLQ} shows the variation of parton-level
total production cross section for $Q\bar Q$ (in red) and
for $Q\chi_1$ and $\bar Q \chi_1 $ (in blue) in $2j + $MET
and in $j+$MET channels, respectively, at the LHC at 13 TeV.
The cross sections presented in Fig.~\ref{fig:VLQ} have been
calculated at leading order using
{\tt MadGraph5}\,\cite{Alwall:2014hca} with the NNPDF2.3LO
parton distribution function\,\cite{Ball:2014uwa}. The
production cross section of $Q\bar Q$ in the $2j + $MET
channel has negligible dependence on $f_{d,s,b}$ since the
dominant parton-level process for the production is
$gg\to Q\bar Q$, which is independent of $f_{d,s,b}$. Hence,
the two red curves, solid for $f_{d,s,b} = 0.1$ and dashed
for $f_{d,s,b} =1$, coincide with each other. However, the
cross section for $Q\chi_1$ and $\bar Q \chi_1 $ in $j+$MET
channels scales as $f_{d,s,b}^2$ since the parton-level
process involved in the production is
$gq,\,g\bar q \to Q\chi_1,\,\bar Q\,\chi_1$, whose amplitude
is proportional to $f_{d,s,b}$. Note that the only possible
decay mode of $Q$ is to a down-type quark and a $\chi_1$. 

To estimate the signature of our model in collider
experiments, events have been generated at partonic level
using {\tt MadGraph5} with the NNPDF2.3LO parton
distribution function using the Universal FeynRules Object
(UFO)\,\cite{Degrande:2011ua} files generated by
{\tt FeynRules}\,\cite{Alloul:2013bka,Christensen:2008py} at
a center-of-mass energy of 13 TeV; partons in the final
state have been showered and hadronized using the parton
shower in {\tt PYTHIA 8.210}\,\cite{Sjostrand:2014zea} with
{\tt 4C} tune\,\cite{Sjostrand:2006za}. Stable particles
have been clustered into anti-kT\,\cite{Cacciari:2008gp}
jets of size 0.4 (used by both ATLAS and CMS) using the
{\tt FastJet}\,\cite{Cacciari:2011ma} software package; only
the jets with $P_{T}$ more than 30 GeV have been considered
for further analysis.

In Fig.~\ref{fig:dist}, we present some important and
representative differential distribution of some observables
as are considered by experimental collaborations to search
for signals. The top-left panel in the figure shows the
distribution of $p_T$ of the leading jet while the panel in
top right shows the distribution for $p_T$ of the second
jet. In the bottom-left panel, we show the distribution of
missing transverse energy ($\slashed p_T$). The bottom-right
panel shows the distribution of
$H_T =\!\!\!\underset{j\,\in\,\text{jets}}{\sum} |\vec p_{T_j}|$,
which is the scalar sum of $p_T$ of all the jets. The major
sources of the SM backgrounds for jets+MET are from the
production of $Z$ decaying to $\nu\bar\nu$ and $W$ decaying
to $\tau\nu_\tau$ in events with jets. Also QCD events are
potential sources to contribute to the same final state. The
distribution for these three backgrounds are plotted in four
panels of Fig.~\ref{fig:dist}. SM background samples have
been generated with at leading order (LO) using
{\tt MadGraph5}\,\cite{Alwall:2014hca} with the NNPDF2.3LO
parton distribution function\,\cite{Ball:2014uwa} at
center-of-mass energy of 13 TeV and
{\tt PYTHIA 8.210}\,\cite{Sjostrand:2014zea}, with the same
{\tt 4C} tune\,\cite{Sjostrand:2006za} as used for
generation of the signal sample, has been used for the
simulation of fragmentation, parton shower, hadronization
and underlying event.  The distribution for QCD, $W$+jets,
and $Z$+jets backgrounds are plotted with gray, purple, and
green, respectively, with the same color convention in all
four panels. From the figure, it is quite clear that the
bumps for signals will not be significant enough to be
observed above the expected fluctuation of the background.

Following the distribution in the experimental
references\,\cite{Aaboud:2017rzf,Aaboud:2017phn,Khachatryan:2014rra,Sirunyan:2017jix,Sirunyan:2017cwe,Sirunyan:2017kqq,Aaboud:2017hrg,Aaboud:2017vwy,Aaboud:2017ayj},
we carried out our analysis with the same distribution. As
discussed earlier, the direct production of new particles
will contribute to $2j + $MET and $j + $MET signals. There
are few dedicated searches in these channels to search for
dark matter
signals\,\cite{Aaboud:2017rzf,Aaboud:2017phn,Khachatryan:2014rra,Sirunyan:2017jix}.
Few other models, especially supersymmetry (SUSY) in the
R-parity conserving scenario, also lead to these kinds of
signals. These searches have also been done by both
CMS\,\cite{Sirunyan:2017cwe,Sirunyan:2017kqq} and
ATLAS\,\cite{Aaboud:2017hrg,Aaboud:2017vwy,Aaboud:2017ayj}.
Though the results are given in terms of SUSY parameters or
effective theory parameters, one can recast the result for a
given model and check for its consistency. But these
searches do not yield any further constraint in the
parameter space in the model. A dedicated search for this
model may give a stronger constraint, but the analysis of
such a search is beyond the scope of this work.

\section{Results}\label{sec:result}

\begin{figure}[h]
\includegraphics[width=0.45\textwidth]{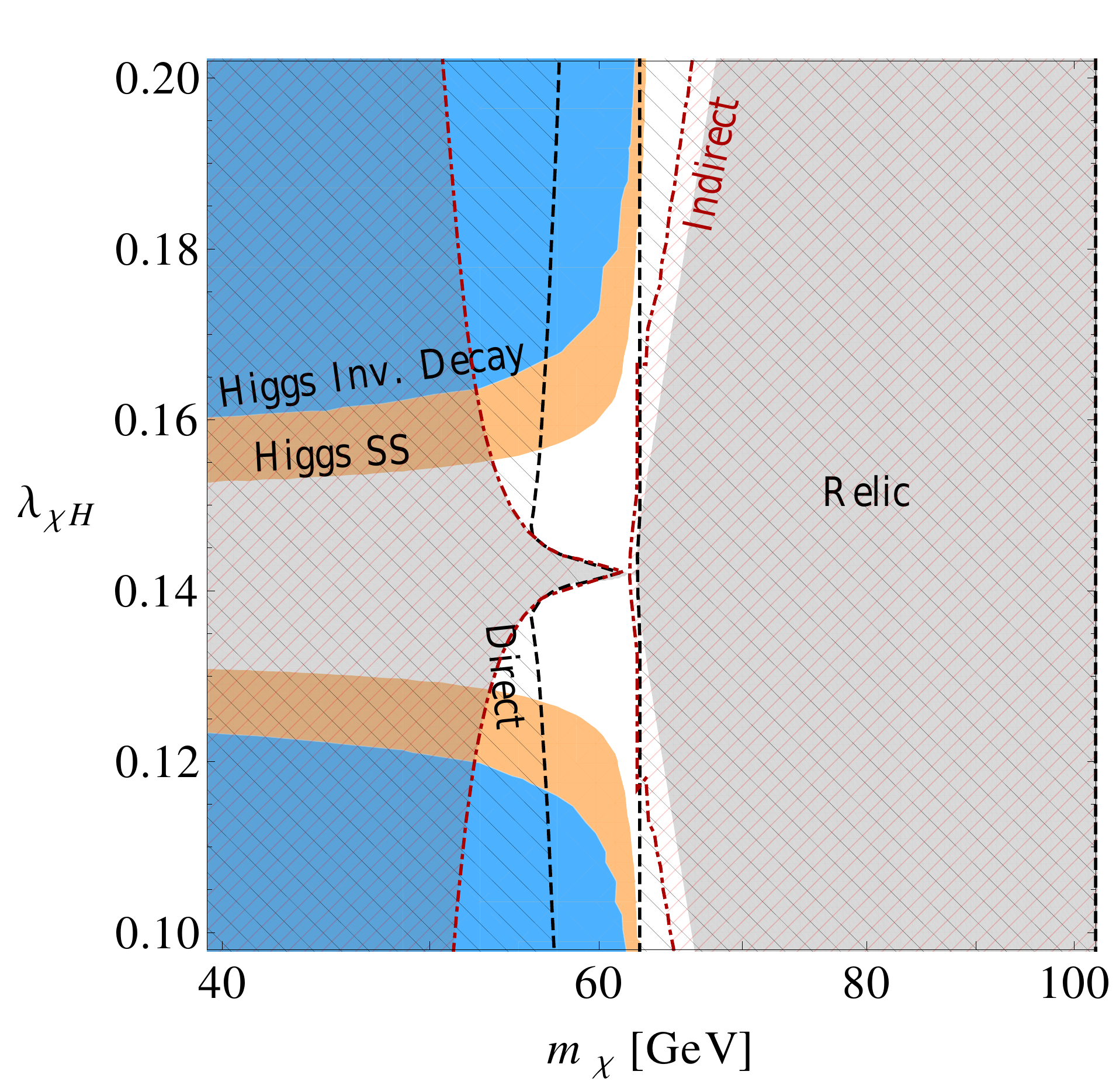}
\caption{Allowed regions in the parameter space for the
two-component axion-WIMP DM model. The gray shaded region
shows the area ruled out by DM relic abundance constraint
corresponding to the $2\sigma$ bound
$\Omega_c h^2<0.12$\,\cite{Ade:2015xua}. The black hatched
lines show the regions of parameter space ruled out by the
DM direct detection bounds from XENON1T$\times$1 yr
experiment\,\cite{Aprile:2018dbl}. The hatched region within
the red curve is ruled out by the DM annihilation data from
DES-Fermi-LAT experiment\,\cite{Fermi-LAT:2016uux}. The blue
shaded region shows the bounds imposed due to the invisible
decay modes of the Higgs, which is roughly $25\%$ of its
branching ratio\,\cite{ATLAS:2018gcr,CMS:2018xuk}. The bound
coming from the signal strength of the Higgs is shown in
orange\,\cite{Aaboud:2017bja,CMS:2018awd}. The white,
unshaded region represents the allowed parameter space in
this model.}\label{fig:final-result}
\end{figure}

Our main results are summarized in
Fig.~\ref{fig:final-result}. The relevant bounds coming from
the different experiments are imposed on the region
satisfying the DM relic density in the
$\lambda_{\chi H}-m_\chi$ plane. The gray shaded region is
ruled out by the relic constraints. We allow for both
$\chi_1$ as well as the axion to contribute to the DM relic
density. Hence the white region, corresponding to the
$2\sigma$ bound $\Omega_c h^2<0.12$, represents the allowed
parameter space, satisfying the relic density. As explained
before, near $m_\chi\approx m_h/2$, the DM annihilation
cross section is enhanced from the Higgs resonance, thereby
decreasing the relic density of DM. This explains why the
allowed region from relic is centered around $m_\chi=m_h/2$.
Furthermore, there is a particular set of parameters for
which $h\chi_1\chi_1$ coupling vanishes, leading to a rise
in the relic density. This accounts for the peaklike
structure in Fig.~\ref{fig:final-result}, which occurs at
$\lambda_{\chi H} \sim 0.14$ for our choice of parameters.

The black hatched lines show the regions of parameter space
ruled out by the direct detection bounds from
XENON1T$\times$1 yr experiment. The hatched region within
the red curve is ruled out by DES-Fermi-LAT joint gamma-ray
search data from the Milky Way satellite galaxies. As is
clearly seen, most of the allowed regions are ruled out,
leaving behind a tiny window around in the
$\mchi-\lambda_{\chi H}$ plane. Clearly, this window is
centered around $\mchi\approx m_h/2$ and the value of
$\lambda_{\chi H}$ for which the $h\chi_1\chi_1$ coupling
vanishes.

The blue shaded region shows the bounds imposed due to the
invisible decay modes of the Higgs, which is roughly $25\%$
of its branching ratio. More stringent bounds are imposed
from the signal strength of the Higgs, which is shown in
orange. These also help to rule out extra regions of the
parameter space for larger as well as smaller values of
$\lambda_{\chi H}$. We have also checked that the LHC bounds
from production of $Q\bar{Q}$ are relatively weak; hence
they do not impose any extra constraint on the model.

Thus, from the above figure, one concludes that only a small
fraction of the model can still be accommodated from
existing experimental bounds. This region, however, enjoys
the advantage of an accidental cancellation of the couplings
near $m_h/2$, thereby making it extremely difficult to rule
out experimentally. This tiny window provides a breathing
space for the model to survive.

\section{Summary and Discussions}\label{sec:conclude}
In this paper, we have performed a comprehensive study of a
two-component dark matter model, consisting of the QCD axion
and an electromagnetic charge neutral scalar particle, both
contributing to the relic density. The theory is symmetric
under a global Peccei-Quinn symmetry, which can be
spontaneously broken down to a residual $\mathbb{Z}_2$
symmetry. For concreteness, we have considered a specific
model: the KSVZ model of the axion, augmented with an
additional complex scalar. After spontaneous breaking of
the PQ symmetry, the residual $\mathbb{Z}_2$ symmetry allows
the lightest component of the complex scalar to be a DM
candidate, apart from the axion. We have tested the model in
the light of recent data from DM direct and indirect search
experiments. Furthermore, we have also studied the different
collider signatures of this model. 

Although the observational and experimental constraints are
found to be very restrictive, a synergy of the enhancement
of DM annihilation from the Higgs resonance and the
vanishing of the coupling between the Higgs and the dark
matter leave room for future experimental investigation of
this model. A large portion of the parameter space predicts
overabundance of $\chi_1$ in the Universe and hence is not
viable. In the remaining underabundant region of $\chi_1$,
the axion can form the dominant part of the CDM. The
viability of the axion being the CDM is being tested in
several ongoing experiments. The latest dark matter direct
and indirect detection experiments results further constrain
this model. Moreover, these results are expected to improve
the bounds by a few orders of magnitudes over the next few
years which will subject this model to even tighter
constraints. Although the bounds from the measurements of
the properties of the Higgs at collider experiments are
relatively weak, they still help to rule out an additional
part of the parameter space. Future measurements of
vectorlike quarks at high-luminosity and high-energy
operating modes of the LHC can shed further light on the
viability of this model.

Higher-order loop corrections may modify the DM direct
detection cross section to some extent, as discussed in
Refs.\,\cite{Klasen:2013btp,Ibarra:2015fqa,Klasen:2016qyz,Abe:2018emu}.
Additionally, virtual internal bremsstrahlung in the
annihilation of $\chi_1$ may introduce special features to
the gamma-ray energy spectrum, making it easier to detect
by some of the experiments\,\cite{Chowdhury:2016bxs}.
However, these corrections were not taken into account here
and will be addressed in a future work. Nevertheless, it is
possible to add new particles to this simplistic model,
e.g., an additional scalar, to enrich its phenomenology and
evade some of the experimental bounds. This leaves room for
future scopes of model building and investigation of
observable signatures in high-energy experiments. In this
work, we have calculated the prediction from our model with
some natural choice for the couplings to compare with
experimental data; but other values of the couplings can be
explored to test the validity of the model on the basis of
available experimental results. In conclusion, the
two-component dark matter model, consisting of the WIMP and
the axion, continues to survive, in spite of being tightly
constrained. 

\begin{acknowledgments}
We thank the Workshop on High Energy Physics Phenomenology
2017 for providing us with an environment for lively and
fruitful discussions where this project started. We thank
Sabyasachi Chakraborty for relevant discussions in the
initial stages of this project and Basudeb Dasgupta for
useful inputs to the project and comments on the manuscript.
We are grateful to Ranjan Laha for pointing out the effects
of higher-order loop corrections in the dark matter direct
detection cross section, as well as the effects of virtual
internal bremsstrahlung in the observed gamma-ray spectrum.
We thank the anonymous referees for the useful suggestions
to improve the manuscript. We acknowledge use of the grid
computing facility in Department of High Energy Physics,
Tata Institute of Fundamental Research, for part of the
Monte Carlo sample generation work. M.\,S. acknowledges
support from the National Science Foundation, Grant No.
PHY-1630782, and to the Heising-Simons Foundation, Grant No.
2017-228. T.\,S. acknowledges financial support from the
Department of Atomic Energy, Government of India, for the
Regional Centre for Accelerator-based Particle Physics
(RECAPP), Harish-Chandra Research Institute.
\end{acknowledgments}

\appendix*
\section{No Symmetry Breaking of $\chi$}
\label{app:chissb}
The potential for $\chi$ before PQ-symmetry breaking is
given by
\begin{equation}
V(\chi) = \lambda_\chi |\chi|^4 + \mu_\chi^2|\chi|^2.
\end{equation}
The minima for this potential occurs at
$v_\chi = \sqrt{-\left(\mu_\chi^2/2\lambda_\chi\right)}$. We
can always choose parameters such that $v_\chi < F_a$. This
will prevent $\chi$ from developing a VEV before PQ-symmetry
breaking. After PQ-symmetry breaking, the potential
governing the evolution of the real components of $\chi$,
viz. $\chi_1$ and $\chi_2$, is given by
\begin{equation}
V\left(\chi_1,\chi_2\right) = \frac{\lambda_\chi}{4}\left(\chi_1^2 + \chi_2^2\right)^2 + \frac{1}{2}\,\mu_{\chi_1}^2\,\chi_1^2 + \frac{1}{2}\,\mu_{\chi_2}^2\,\chi_2^2,\label{eqn:pot-chi1chi2}
\end{equation}
where
\begin{eqnarray}
\mu_{\chi_1}^2 = \frac{1}{2}\left(2\mu_\chi^2 + \lambda_{\zeta\chi}F_a^2 - 2\sqrt{2}\epsilon_\chi F_a\right),\label{eqn:muchi1sq}\\
\mu_{\chi_2}^2 = \frac{1}{2}\left(2\mu_\chi^2 + \lambda_{\zeta\chi}F_a^2 + 2\sqrt{2}\epsilon_\chi F_a\right).
\end{eqnarray}
There are four possibilities depending on the nature of the
parameters:
\begin{enumerate}
    \item[(i)] $\mu_{\chi_1}^2,\,\mu_{\chi_2}^2 > 0$.---This
    does not lead to a vev for $\chi_1$ or $\chi_2$.

    \item[(ii)] $\mu_{\chi_1}^2>0$ and
    $\mu_{\chi_2}^2<0$.---This scenario is not possible
    since we choose $\epsilon_\chi > 0 $ in our analysis.

    \item[(iii)] $\mu_{\chi_1}^2<0$ and
    $\mu_{\chi_2}^2>0$.---This parameter choice ensures no
    VEV for $\chi_2$. The minimization condition for
    $\chi_1$ then becomes
    \begin{eqnarray}
    & &\frac{\partial V(\chi_1,\chi_2)}{\partial \chi_1}\,\Bigg|_{\chi_1=\chionevev,\,\chi_2=0}
    = \lambda_\chi \chionevev^3 + \mu_{\chi_1}^2\chionevev= 0\,,\nonumber\\
    \end{eqnarray} 
    which leads to
    $\chionevev = 0, {\rm \ and \ } \chionevev = \pm\sqrt{-\left(\mu_{\chi_1}^2/\lambda_\chi\right)}$.
    Clearly, $\chionevev = 0$ is the solution for the
    maxima, and the other two solutions correspond to the
    minima.

    \item[(iv)]
    $\mu_{\chi_1}^2,\,\mu_{\chi_2}^2 < 0$.---This scenario
    is more involved and requires minimization with respect
    to both the fields:
    \begin{eqnarray}
    & &\frac{\partial V(\chi_1,\chi_2)}{\partial \chi_1}\,\Bigg|_{\chi_1=\chionevev,\,\chi_2=\chitwovev} \nonumber\\
    & &\quad = \left[\lambda_\chi \left(\chionevev^2 + \chitwovev^2 \right) + \mu_{\chi_1}^2\right]\chionevev= 0,\\
    \& && \frac{\partial V(\chi_1,\chi_2)}{\partial \chi_2}\,\Bigg|_{\chi_1=\chionevev,\,\chi_2=\chitwovev} \nonumber\\
    & &\quad = \left[\lambda_\chi \left(\chionevev^2 + \chitwovev^2 \right) + \mu_{\chi_2}^2\right]\chitwovev= 0.
    \end{eqnarray}
    The above set of equations permits the following
    solutions:
    \begin{eqnarray}
    ({\rm soln:a}) &&\quad \chionevev = \chitwovev = 0, \label{eqn:sola}\\
    ({\rm soln:b}) &&\quad \chionevev = 0,\ \chitwovev = \pm\sqrt{-\frac{\mu_{\chi_2}^2}{\lambda_\chi}}, \label{eqn:solb}
    \end{eqnarray}
    \begin{eqnarray}
    {\rm and\ } ({\rm soln:c}) \quad \chionevev=\pm\sqrt{-\frac{\mu_{\chi_1}^2}{\lambda_\chi}},\ \chitwovev=0.~~ \label{eqn:solc}
    \end{eqnarray}
    To determine which solution corresponds to a minima, we
    take a further derivative to get
    \begin{eqnarray}
    && \frac{\partial^2 V(\chi_1,\chi_2)}{\partial\chi_1^2} = \lambda_\chi\left(3\chi_1^2 + \chi_2^2\right) + \mu_{\chi_1}^2,\\
    && \frac{\partial^2 V(\chi_1,\chi_2)}{\partial\chi_2^2} = \lambda_\chi\left(\chi_1^2 + 3\chi_2^2\right) + \mu_{\chi_2}^2,\\
    && \frac{\partial^2 V(\chi_1,\chi_2)}{\partial\chi_1\partial\chi_2} = 2\,\lambda_\chi\,\chi_1\,\chi_2.
    \end{eqnarray}
    The nature of the solutions in Eqs.~(\ref{eqn:sola},
    \ref{eqn:solb}, and \ref{eqn:solc}) are determined by
    the determinant and trace of the Hessian. This gives the
    following conditions:
    \begin{itemize}
        \item[$({\rm soln:a})$] Det = $0$ and Tr = $0$.---Further
        analysis is required to tell the nature of this
        point. For this solution, $V(\chi_1=0,\chi_2=0) =0$,
        which is bigger than the values at other solutions.
        So, even if it is a minima, it is not a global
        minima.

        \item[$({\rm soln:b})$] Det =
        $4\sqrt{2}\epsilon_\chi F_a \mu_{\chi_2}^2 < 0$.---This
        corresponds to a saddle point solution.

        \item[$({\rm soln:c})$] Det =
        $-4\sqrt{2}\epsilon_\chi F_a \mu_{\chi_1}^2 > 0$ and
        Tr = $\mu_{\chi_2}^2-3\mu_{\chi_1}^2 > 0$.---This is
        a minima.
    \end{itemize}
\end{enumerate}

\begin{figure}[h]
\includegraphics[width=0.42\textwidth]{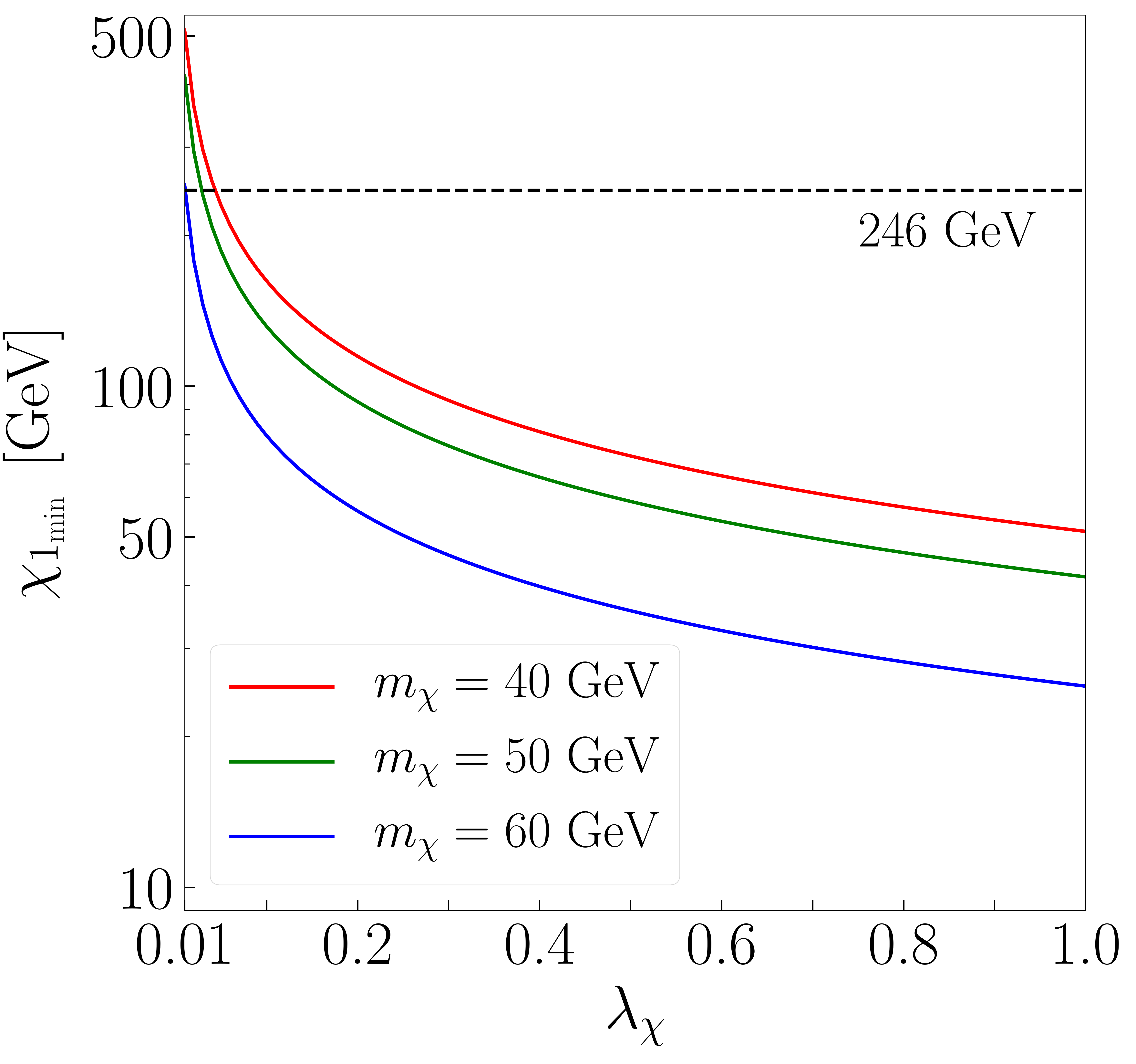}
\caption{Values of $\chionevev$ for our parameter choices as
a function of $\lambda_\chi$ for different values of
$m_\chi$.}\label{fig:ch1vevvslchi}
\end{figure}

This analysis confirms that $\chi_2$ will never get a VEV.
Whether $\chi_1$ will get a VEV or not depends on the choice
of the parameters. For our model to be valid, we need to
choose parameters in such a way so that $\chionevev < v_H$.
This will prevent $\chi_1$ from developing a VEV before
electroweak symmetry breaking. After electroweak symmetry
breaking, however, $\chi_1$ gets real mass as given in
Eq.~(\ref{masschi}). We can quickly estimate the value of
$\chionevev$ for the parameters of our model. The values of
$\chionevev$ as a function of $\lambda_\chi$ are shown in
Fig.~\ref{fig:ch1vevvslchi} for different values of
$m_\chi$. Clearly, for natural values of the coupling
$\lambda_\chi \gtrsim 0.05$, global minima of the potential
in Eq.~(\ref{eqn:pot-chi1chi2}) occurs for values
$\chionevev$ below $v_H$, which prevents $\chi_1$ from
developing a VEV before electroweak symmetry breaking. After
electroweak symmetry breaking, both $\chi_1$ and $\chi_2$
get real mass. So, we can always choose parameters in our
model in such a way that neither $\chi$ nor $\chi_1$ or
$\chi_2$ develops a VEV throughout the entire
symmetry-breaking process of $\zeta$ and $H$.

%

\end{document}